\documentclass[12pt]{article}

\usepackage{array,epsfig}
\usepackage{float}
\usepackage{amsmath}
\usepackage{amsfonts}
\usepackage{amssymb}
\usepackage{amsxtra}
\usepackage{amsthm}
\usepackage{dsfont}
\usepackage{mathrsfs}
\usepackage{pgf,tikz}
\usepackage{mathrsfs}
\usepackage{slashed}
\usepackage{mathtools}
\usepackage{color}
\usepackage{braket}
\usepackage{listings}
\usepackage{caption}
\usepackage{subcaption}
\usepackage{hyperref}
\usepackage[title]{appendix}
\usepackage{graphicx}
\usetikzlibrary{calc}
\usetikzlibrary{arrows}
\usetikzlibrary{plotmarks}

\usepackage[backend=biber,sorting=none,style=ieee,citestyle=numeric-comp,isbn=false,dashed=false]{biblatex}
\usepackage{bm}
\usepackage{textcomp}
\usepackage{gensymb}

\theoremstyle{definition}

\graphicspath{{figures/}}

\definecolor{arxivc}{RGB}{22,72,145}
\def\arxivtourl#1{http://arxiv.org/abs/#1}      
\DeclareFieldFormat{eprint:arxiv}{%
    \ifhyperref
        {\color{arxivc}\href{\arxivtourl#1}{[arxiv:\nolinkurl{#1}]}}
        {\nolinkurl{#1}}}

\newcommand{\dbar}{\bar{d}}

\newcommand{\optwo}{O_2}

\newcommand{\targn}{\tilde{N}_{31}}

\newcommand{\vthet}{\vec{\theta}}
\newcommand{\vthetm}{\vec{\theta}_\textrm{min}}

\let\vec\boldsymbol

\begin{document}

\begin{center}
${}$\\
\vspace{100pt}
{ \Large\bf On the Nature of Spatial Universes \\
\vspace{15pt}\, in 3D Lorentzian Quantum Gravity
}

\vspace{46pt}

{\sl J. Brunekreef}
and {\sl R. Loll}

\vspace{24pt}
{\footnotesize

Institute for Mathematics, Astrophysics and Particle Physics, Radboud University \\ 
Heyendaalseweg 135, 6525 AJ Nijmegen, The Netherlands.\\ 
\vspace{12pt}
{email: jorenb@gmail.com, r.loll@science.ru.nl}\\
}
\vspace{24pt}

\end{center}

\vspace{0.8cm}

\begin{center}
{\bf Abstract}
\end{center}

\noindent Three-dimensional Lorentzian quantum gravity, expressed as the continuum limit of a nonperturbative sum over spacetimes, 
is tantalizingly close to being amenable to analytical methods, and some of its properties have been described in terms of effective matrix and other
models. To gain a more detailed understanding of three-dimensional quantum gravity, we perform a numerical
investigation of the nature of spatial hypersurfaces in three-dimensional Causal Dynamical Triangulations (CDT). We measure 
and analyze several quantum observables, the entropy exponent, the local and global Hausdorff dimensions, and the quantum Ricci 
curvature of the spatial slices, and try to match them with known continuum properties of systems of two-dimensional quantum geometry. 
Above the first-order phase transition of CDT quantum gravity, we find strong evidence that the 
spatial dynamics lies in the same universality class as two-dimensional Euclidean (Liouville) quantum gravity. Below the transition, 
the behaviour of the spatial slices does not match that of any known quantum gravity model. This may indicate the existence of a new
type of two-dimensional quantum system, induced by the more complex nature of the embedding three-dimensional quantum geometry.   

\vspace{12pt}
\noindent

\newpage

\tableofcontents

\section{Introduction}
A complete theory of quantum gravity may offer insights into how the spacetime we observe and inhabit can emerge from first principles. 
The nonperturbative gravitational path integral is a promising route toward such a theory, formulated within a purely
quantum field-theoretic setting \cite{loll2022questions}.
If one is interested in concrete Planckian or near-Planckian results in the full, four-dimensional theory, like information on 
the spectra of diffeomorphism-invariant observables, 
Causal Dynamical Triangulations or {\it CDT quantum gravity} \cite{ambjorn2012nonperturbative,loll2019quantum} is arguably the path integral 
approach that is furthest developed. Recall that the continuum path integral for pure gravity is given by
\begin{equation}
Z=\int\limits_{{\cal G}(M)}\!\! {\cal D} [g]\, {\rm e}^{\, i S^{\rm EH}[g]},\;\;\;\;\;\;
S^{\rm EH}[g]=\frac{1}{16\pi G_{\rm N}}\, \int\limits_M d^4 x\, \sqrt{ -\det(g)}\, (R -2 \Lambda ),
\label{pathint}
\end{equation}
where ${\cal G}(M)$ denotes the space of diffeomorphism-equivalence classes $[g]$ of Lorentzian metrics $g_{\mu\nu}(x)$ on the manifold $M$,
and $S^{\rm EH}$ is the Einstein-Hilbert action. In the CDT set-up this formal expression 
is given a precise meaning, namely, as the continuum limit of a regularized version of (\ref{pathint}), 
with ${\cal G}(M)$ approximated by a 
space of piecewise flat Lorentzian spacetimes. Although the primary physical interest is in spacetime dimension $D\! =\! 4$, 
the CDT path integral has also been studied in two and three dimensions. 

Hallmarks of this strictly nonperturbative approach are (i) the presence of a well-defined analytic continuation or ``Wick rotation'', mapping 
the regularized 
path integral to a real partition function, which enables its analytical evaluation in $D\! =\! 2$ \cite{ambjorn1998nonperturbative} and numerical
evaluation in $D\! =\! 2$, 3 and 4 \cite{ambjorn2000nonperturbative,ambjorn2001dynamically}, (ii) its formulation on a space of {\it geometries}, 
avoiding the need to gauge-fix the diffeomorphism symmetry and isolate its physical
degrees of freedom, (iii) following the logic of critical phenomena, a high degree of uniqueness and universality if a continuum limit can be
shown to exist, (iv) a nonperturbative cure of the conformal divergence, which by default renders Euclidean path integrals in $D\!\geq\! 3$ ill defined \cite{dasgupta2001propertime}, and (v) unitarity, in the form of reflection posi\-tivity of the regularized path integral, with respect to a notion of discrete proper time \cite{ambjorn2001dynamically,ambjorn2012nonperturbative}. 

In terms of results in $D\! =\! 4$, in addition to the presence of second-order phase transitions \cite{ambjorn2011secondorder,ambjorn2012second,coumbe2016exploring}, 
necessary for the existence of a continuum limit,
an important finding of CDT is the emergence of an extended four-dimensional universe \cite{ambjorn2004emergence,ambjorn2021cdt}. 
With the standard choice $M\! =\! S^1\!\times\! S^3$ for the topology, in terms of the quantum observables measured so far 
(spectral and Hausdorff dimensions \cite{ambjorn2005spectral,ambjorn2005reconstructing}, shape of the universe, including quantum 
fluctuations \cite{ambjorn2008planckian,ambjorn2008nonperturbative}, average Ricci curvature \cite{klitgaard2020how}), 
its behaviour on sufficiently coarse-grained scales is compatible with that of a de Sitter universe.
This is remarkable because it represents nontrivial evidence of a classical limit,
one of the high hurdles to clear for any nonperturbative and manifestly background-independent approach to quantum gravity.

After Wick rotation, the gravitational path integrals of CDT become partition functions of  
statistical systems, whose elementary geometric building blocks (flat $D$-dimensional simplices, see Sec.\ \ref{sec:phases} for further details) 
are assembled into piecewise flat manifolds $T$ -- the triangulations --
each one contributing with a Boltzmann weight $\exp (-S^{\rm EH}[T])$.\footnote{Note that $S^{\rm EH}[T]$ is the so-called bare action of the
regularized theory, depending on bare coupling constants, which in the continuum limit will typically undergo renormalization.} 
There are a couple of reasons why such seemingly simple ingredients can give rise to interesting continuum theories of quantum gravity and quantum
geometry. On the one hand, there is the highly nontrivial combinatorics of how the simplicial building blocks can be glued together to yield 
distinct curved spacetimes $T$. Especially in dimension $D\!\geq\! 3$, this reflects the complexities of local geometry and curvature, already familiar 
from the classical theory. On the other hand, there is
a complicated interplay between ``energy'' (the bare action) and ``entropy'' (the number of distinct triangulations for a given value of the bare action),
which depends on the values of the bare coupling constants, i.e.\ the point in phase space at which the path integral is evaluated. 

An enormous amount has been learned about such nonperturbative systems of geometry over the last 35 years, beginning with the Euclidean
analogue and precursor of the Lorentzian CDT theory, based on Euclidean dynamical triangulations or dynamical triangulations (DT) for short 
\cite{david1985planar,kazakov1985critical,ambjorn1997quantumb}. A crucial role in the exploration of these systems 
has been played by Monte Carlo methods, which are employed to numerically evaluate the path integral and expectation values of
observables by importance sampling \cite{binder1997applications,newman1999monte}. This is also true in dimension $D\! =\! 2$,
where in addition a variety of nonperturbative analytical solution techniques are available, e.g.\ combinatorial, matrix model and transfer matrix methods \cite{DiFrancesco:1995ih,ambjorn1998nonperturbative,ambjorn1999euclidean}, leading to compatible results.

Monte Carlo simulations should be seen as numerical experiments, providing tests and feedback for the construction of the 
theory. For full quantum gravity, the quantitative information on the nonperturbative sector obtained from numerical analysis 
is extremely valuable, since it cannot currently be 
substituted by anything else. Although we do not know in detail what a theory of quantum gravity will eventually look like, it seems unlikely that
it will come in closed analytic form. A potential scenario would be akin to QCD, where we manage to
extract nonperturbative information about the theory's spectrum (of suitable quantum-geometric observables) with ever greater accuracy,
using a background-independent analogue of lattice gauge theory such as (C)DT. 
Despite its conventional, quantum field-theoretic setting and the absence of any exotic ingredients, this type of lattice gravity has already uncovered 
unexpected features of strongly quantum-fluctuating geometry, like the dynamical dimensional reduction of 
spacetime near the Planck scale \cite{ambjorn2005spectral}, which is conjectured to be universal \cite{carlip2017dimension}. 

The focus of the present work will be the Lorentzian CDT path integral in {\it three} spacetime dimensions \cite{ambjorn2001nonperturbative}.
More specifically, as a stepping stone towards a more detailed geometric understanding of this quantum gravity model, we will
investigate the geometry of its two-dimensional spatial hypersurfaces. A key question is whether in
a continuum limit the behaviour of these surfaces falls into one of the known universality classes \cite{goldenfeld2019lectures} 
of nonperturbative quantum gravity in two dimensions, or whether there is evidence for a different type of quantum dynamics.  
The two universality classes in question are that of (the scaling limit of) two-dimensional DT \cite{david1985planar,ambjorn1997quantumb}, 
which also contains Liouville quantum gravity, and that of two-dimensional CDT quantum gravity \cite{ambjorn1998nonperturbative,ambjorn2013universality}.    

Our study will be numerical in nature, but -- depending on the outcome -- may well provide input for further analytical work, which could be 
technically feasible because of the effective two-dimensional character of the spatial slices. Note
that we do not claim that there is a direct physical interpretation of the properties of these spatial geometries from a three-dimensional point of view
(inasmuch as a lower-dimensional toy model of quantum gravity can be called ``physical'' at all).
Although in our set-up a spatial slice at constant proper time is an invariantly defined concept\footnote{It is defined as the set of all points at a given proper-time 
distance to a given initial spatial surface or an initial singularity, in the spirit of similar constructions in the continuum \cite{andersson1998cosmological}. 
Note also that the proper-time slicing is not related to any gauge-fixing, since the CDT set-up is manifestly diffeomorphism-invariant 
(see e.g.\ \cite{loll2019quantum} for a detailed discussion).}, 
it is not clear to what extent its properties can be thought of as ``observable'', because of the highly nonlocal construction of the hypersurfaces and because 
of their singular nature (``moments in time'') from the point of view of the quantum theory. 
To obtain true quantum observables in a three-dimensional, spacetime sense would presumably require some smearing in the time direction. 
 
Nevertheless, our measurements within the slices of constant time are perfectly well defined operationally and 
give us a quantitative handle on the influence of the three-dimensional quantum geometry in which the spatial slices are embedded, as we will 
demonstrate. We will start by investigating the distribution of the vertex order in the slices, which counts the number of spatial edges meeting
at a vertex. This quantity is not per se related to a continuum observable, but can be compared with known exact results for the 
ensembles of two-dimensional DT and CDT geometries.
The core of the paper consists of measuring and analyzing the following quantum observables: (i) the entropy exponent $\gamma$, also known as the string susceptibility, which determines the subexponential growth
of the partition function at fixed two-volume $A$, as a function of $A$; (ii) the Hausdorff dimension $d_H$, obtained by
comparing volumes with their linear extension, where we distinguish between a local and
a global variant; (iii) the so-called curvature profile $\mathcal{R}(\delta)$ of the spatial slices, measuring the average 
quantum Ricci curvature \cite{klitgaard2018introducing} of the surfaces as a function of a linear coarse-graining scale $\delta$.
We find convincing evidence that the effective dynamics of the spatial slices in the so-called degenerate phase of three-dimensional CDT
quantum gravity is described by two-dimensional DT quantum gravity. However, we do not find a match with any known two-dimensional
system of quantum geometry in the so-called de Sitter phase, where the dynamics of the hypersurfaces is much richer due to the nontrivial
influence of the embedding three-geometry. Further research is needed to determine the continuum nature of the effective spatial dynamics in
this phase.

The remainder of the paper is structured as follows. 
In the next section, we recall the main ingredients of CDT quantum gravity in $D\! =\! 3$ and review previous research on the subject,
and what it has revealed about its phase structure and physical characteristics. 
In Sec.\ \ref{sec:impl}, we discuss the numerical implementation of the three-dimensional CDT path integral in terms of Markov chain Monte Carlo methods. 
Sec.\ \ref{sec:2d-qg} contains a detailed description of the properties of the spatial slices that we have studied numerically. 
We present the results of our measurements, and describe the overall picture that emerges from them. 
In Sec. \ref{sec:disc} we summarize and discuss our findings.
A couple of technical discussions have been relegated to appendices, to improve the readability of the main part of the paper.

\section{Three-dimensional CDT quantum gravity}
\label{sec:phases}

Quantum gravity in three spacetime dimensions \cite{carlip1998quantum} provides an interesting test case for the full gravitational path integral. 
Although the pure gravity theory does not have any local propagating degrees of freedom, the path integral
has the same functional form in terms of the three-dimensional metric as its four-dimensional counterpart (\ref{pathint}), and therefore looks equally 
ill-behaved with regard to its behaviour under renormalization.
How to reconcile the difficulties of solving this metric path integral with the ``topological'' nature of three-dimensional gravity\footnote{More 
precisely, the physical degrees of freedom of three-dimensional gravity are global modes of the metric, described by Teichm\"uller parameters, 
which are present when the genus of the spatial slices is larger than or equal to 1. The present work uses spherical slices, without such parameters.}, 
which leads to considerable simplifications in a first-order, Chern-Simons formulation, without any quantum field-theoretic divergences, 
is only partially understood (see \cite{carlip2005quantum} for a discussion). 

The CDT formulation has thrown some light on the nonperturbative aspects of this question, uncovering both similarities and 
differences between the three-dimensional and the physical, 
four-dimensional theory \cite{ambjorn2001nonperturbative,ambjorn2001computer,ambjorn20023d}. 
For a better understanding of the issues involved and to set the stage for the main part of the paper, let us briefly recall the set-up in three dimensions.
After applying the Wick rotation mentioned in the previous section, the regularized CDT path integral in $D\! =\! 3$ takes the form of a partition function
\begin{equation}
Z = \sum\limits_{\text{triang.}\, T}\frac{1}{C_T}\, {\rm e}^{-S^{\rm EH}[T]}, \;\;\;\;\;\;\;
S^{\rm EH}[T]=-k_0 N_0(T)+k_3 N_3 (T),
\label{cdtpi}
\end{equation}
where $S^{\rm EH}[T]$ denotes the Regge form of the Einstein-Hilbert action on the piecewise flat triangulation $T$, $N_0(T)$ and $N_3(T)$
are the numbers of vertices (``zero-simplices'') and tetrahedra (``three-simplices''), and $C_T$ is the order of the automorphism group of $T$. 
The coupling $k_0$ is proportional to the inverse bare Newton constant and $k_3$ depends linearly on the bare cosmological constant (see
\cite{ambjorn2001nonperturbative} for details). The sum is taken over simplicial manifolds
of a given, fixed topology, which in our case will be $S^1\!\times\! S^2$, a periodically identified time interval times a two-sphere.

\begin{figure}[t]
	\centering
	\includegraphics[width=0.6\textwidth]{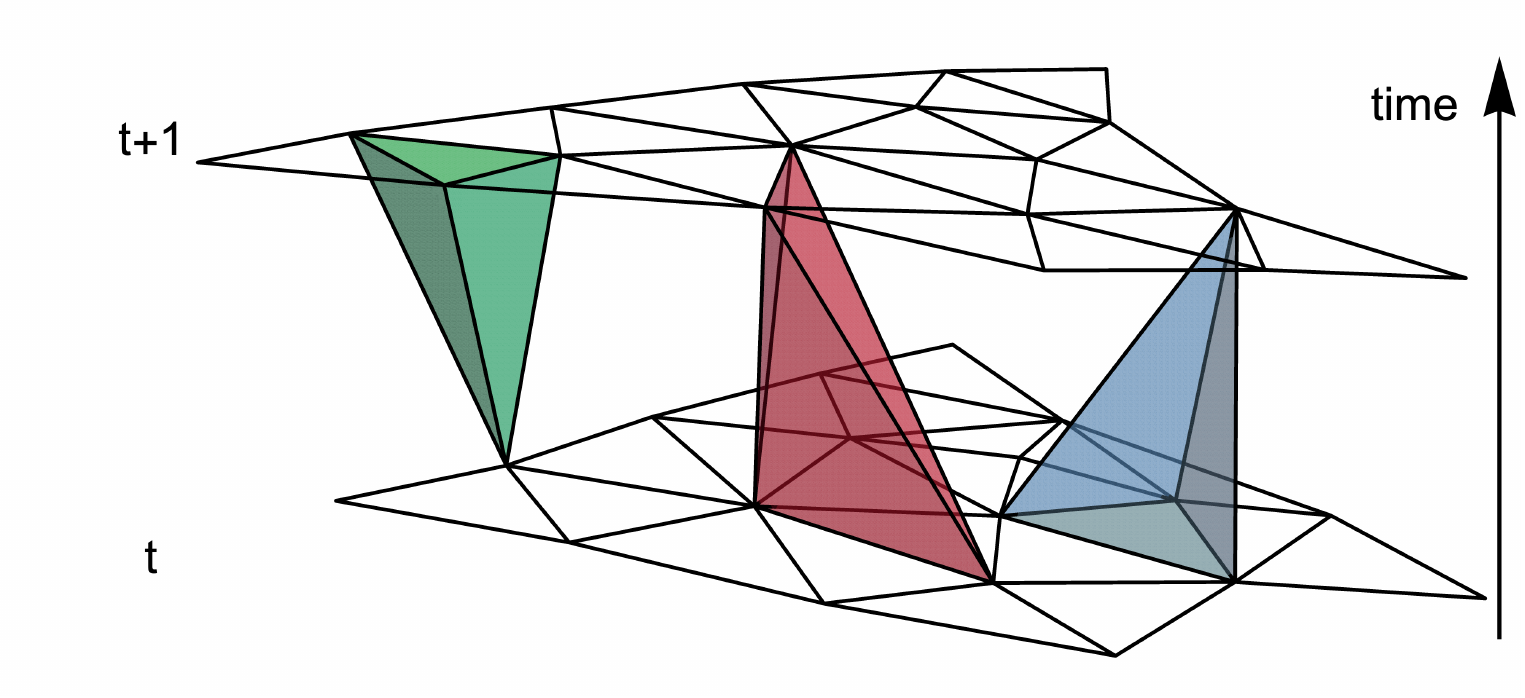}
	\caption{Three types of tetrahedral building blocks of three-dimensional CDT: type (1,3)  (left), type (2,2) (centre), 
	and type (3,1) (right). Note that the two-dimensional triangulations at times $t$ and $t\! +\! 1$ are not drawn isometrically; 
	they are in general curved surfaces.}
	\label{fig:simplices}
\end{figure}

The triangulated configurations $T$ of the Lorentzian CDT path integral have a discrete product structure, representing a simplicial
version of global hyperbolicity, and are assembled from flat, Minkowskian tetrahedra \cite{ambjorn2012nonperturbative,loll2019quantum}. 
A given spacetime geometry can be thought of as a sequence 
of two-dimensional curved, spacelike triangulations, labeled by an integer proper time $t=1,2,3,\dots, t_{\rm tot}$, and made of equilateral triangles. 
The spacetime volume between each pair of adjacent constant-time slices is completely filled in with tetrahedra,
resulting in a ``sandwich'' of simplicial three-dimensional spacetime with topology $[0,1] \times S^2$. The tetrahedral edges linking 
neighbouring spatial slices are \emph{timelike} (and all of equal length), while the edges lying within a spatial slice are of 
course \emph{spacelike} (and also of equal length).
We can therefore classify the building blocks according to their constituent vertices. A tetrahedron of type $(p,q)$ is defined as having
$p$ vertices in slice $t$ and $q$ vertices in slice $t\! +\! 1$,
giving rise to the types (1,3), (2,2) and (3,1), as illustrated by Fig.\ \ref{fig:simplices}. Note that up to time reversal a (1,3)- and a (3,1)-tetrahedron are
geometrically identical.
Since the analytic continuation only affects the edge length assignments and not the topology of the triangulation, this 
characterization of the tetrahedra continues to be meaningful after the Wick rotation.

The key findings of the original, mostly numerical investigation of three-dimensional CDT quantum gravity on $S^1\!\times\! S^2$ inside 
the range $k_0\!\in\! [3,7]$ were as follows \cite{ambjorn2001nonperturbative,ambjorn2001computer,ambjorn20023d}. 
After fine-tuning the bare ``cosmological'' constant $k_3$ to its critical value from inside the region of convergence\footnote{This region
exists because the number of three-dimensional CDT configurations is exponentially bounded as a function of the discrete 
volume $N_3$ \cite{durhuus2015exponential}.} of the partition function Z, 
a two-phase structure was found, consisting of what we shall call a \emph{degenerate phase} for $k_0\!\geq k_0^{\rm c}$ 
and a \emph{de Sitter phase} for $k_0\!\leq k_0^{\rm c}$. These two phases are very reminiscent of corresponding phases in
four-dimensional CDT quantum gravity \cite{ambjorn2012nonperturbative} with regard to their volume profiles, i.e.\ the behaviour of
their spatial volume $V_2$ as a function of the proper time $t$. In the degenerate phase, $V_2(t)$ oscillates wildly, indicating
that spacetime disintegrates into a sequence of uncorrelated two-dimensional geometries (see also Fig.\ \ref{fig:volprofs} below). 
By contrast, in the de Sitter phase a nontrivial 
``blob'' forms when the time extension is chosen sufficiently large, whose shape matches that of a Euclidean de Sitter space (with
$\langle V_2(t)\rangle \propto \cos^2 (\mathit{c}\, t)$), analogous to what has been observed in 
$D\! =\! 4$.\footnote{Volume profiles for nonperiodic boundary conditions in time, with random, fixed two-spheres of various
sizes as spatial boundaries
have been considered in \cite{cooperman2014first,cooperman2017second}.} 
This constitutes nontrivial evidence that a well-defined and macroscopically three-dimensional ground state of geometry\footnote{in the
sense of minimizing the effective Euclidean action governing the nonperturbative quantum dynamics} exists nonperturbatively.
However, unlike in four dimensions the transition at the critical point $k_0^c$ appears to be a first- and not a second-order phase transition,
and no fine-tuning of the inverse gravitational coupling $k_0$ is needed to obtain a continuum limit. This is 
in line with the expectation that no higher-order transitions are present, due to the absence of propagating degrees of freedom. 

Following these results, three-dimensional CDT quantum gravity has been studied from various perspectives. 
Considerable effort has been focused on the transfer matrix associated with a single time step $\Delta t\! =\! 1$, which 
captures the amplitude of going from one spatial two-geometry to an adjacent one. More precisely, one usually considers a simpler, reduced
transfer matrix, whose in- and out-states are labelled by the spatial two-volume (and possibly Teich\-m\"uller parameters). Given our knowledge about
the physical degrees of freedom of three-dimensional gravity, these are the parameters
that are \emph{expected} to be the only relevant ones in the continuum limit. 
The sandwich geometries contributing to the transfer matrix are closer to
two-dimensional quantities and therefore potentially more amenable to an analytic treatment.
In this spirit, a variant of the model was introduced in \cite{ambjorn2001lorentzian}, in which the (1,3)- and (3,1)-building blocks are 
substituted by (1,4)- and (4,1)-pyramids, something that is not expected to affect the universal properties of the model.  
The motivation for considering this variant is that taking a midsection of a sandwich geometry at half-integer time yields a quadrangulation, whose dual
graph is a configuration described by a Hermitian two-matrix model with $ABAB$-interaction, for which analytical results are available. 
However, the bicoloured graph configurations generated by the matrix model form a much larger class than those coming from CDT sandwich geometries,
and correspond to geometries that in general violate the simplicial manifold conditions of the two-dimensional slices and of the 
interpolating three-dimensional piecewise flat geometries in specific ways. Three of these four conditions (following the enumeration in 
\cite{ambjorn2001lorentzian}) are considered mild, in the sense that 
violating them is conjectured not to affect the universality class of the CDT model. As corroborating evidence, \cite{ambjorn2001lorentzian} cites
new numerical simulations of the CDT model in $D\! =\! 3$, where these conditions are relaxed, but which
nevertheless reproduce
the results found in the de Sitter phase of the earlier work that used strict simplicial manifolds \cite{ambjorn2001nonperturbative}. 
Interestingly, they note that the degenerate phase completely disappears in the simulations of this generalized variant of CDT quantum gravity,
and conjecture that the presence of this phase constitutes a discretization artefact. Although our present work does not directly address
these various conjectures (and works with simplicial manifolds only), 
our results suggest that there may be more scope for different universality classes in three dimensions than has been considered up to now, and that it
may be fruitful to re-examine the influence of regularity conditions on continuum results in greater detail. (Of course, not all universality classes
may be associated with interesting models of quantum gravity.) A similar sentiment was expressed in
\cite{durhuus2020structure}, which gives a precise characterization of the bicoloured two-dimensional cell complexes associated with midsections
of CDT geometries with spherical and disc-like spatial slices.

The configurations described by the $ABAB$-matrix model violate also a fourth regularity condition \cite{ambjorn2001lorentzian}, which is associated with a
considerable enlargement of the space of three-geometries. It allows for the appearance of spatial wormholes and a new phase, not present
in standard CDT quantum gravity, where these wormholes are abundant. Whether this phase is interesting from a physics point of view remains
to be understood. The association of CDT quantum gravity with the $ABAB$-matrix model was also used to
analyze the behaviour of the bare coupling constants of the former under renormalization \cite{ambjorn2004renormalization}.
An asymmetric version of the matrix model was studied in
\cite{ambjorn20033d}, motivated by the search for a Hamiltonian associated with the reduced transfer matrix.  
Without invoking matrix models, a continuum Hamiltonian of this kind was derived for the first time in \cite{benedetti2007dimensional}, for spatial slices of cylinder topology,
albeit for a sub-ensemble of CDT configurations with certain ordering restrictions. 
The effective action for the two-volumes of spatial slices with toroidal topology was investigated in \cite{budd2013exploring}, and the dynamics of its Teichm\"uller parameters in \cite{budd2012thesis,budd2012effective}. The phase structure of a one-dimensional balls-in-boxes model, meant to
capture the effective dynamics of the two-volume of CDT quantum gravity, was analyzed in \cite{benedetti2017capturing}, and shown to reproduce
certain features of CDT as well as Ho\v rava-Lifshitz-inspired gravity models (see also \cite{benedetti2015spacetime}).
The spectral dimension of the CDT model in the de Sitter phase was measured and found to be compatible with the classical value of 3 on
large scales and to exhibit a dynamical dimensional reduction to a value compatible with 2 on short scales, similar to what happens in CDT for 
$D\! =\! 4$ \cite{benedetti2009spectral}. Lastly, a generalized model of CDT quantum gravity with causally well-behaved configurations, but without
a preferred proper-time slicing was defined and investigated numerically, and found to reproduce the volume profile of a de 
Sitter space \cite{jordan2013causal,jordan2013sitter}.

\section{Implementation}
\label{sec:impl}

\begin{figure}[t]
    \centering
    \includegraphics[width=0.40\textwidth]{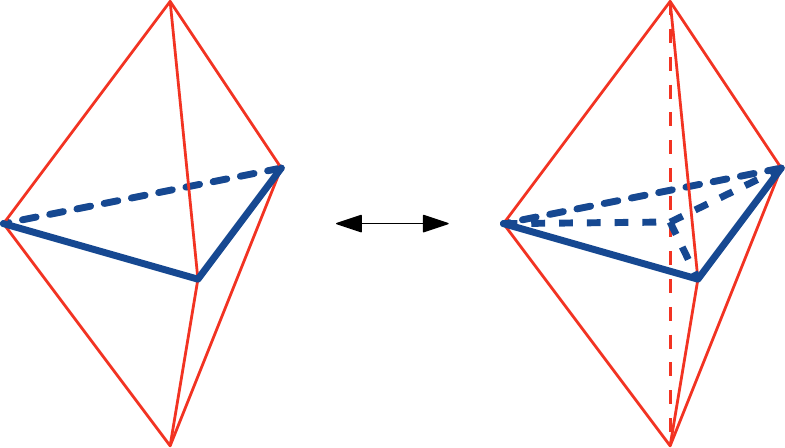} \hfill
    \includegraphics[width=0.40\textwidth]{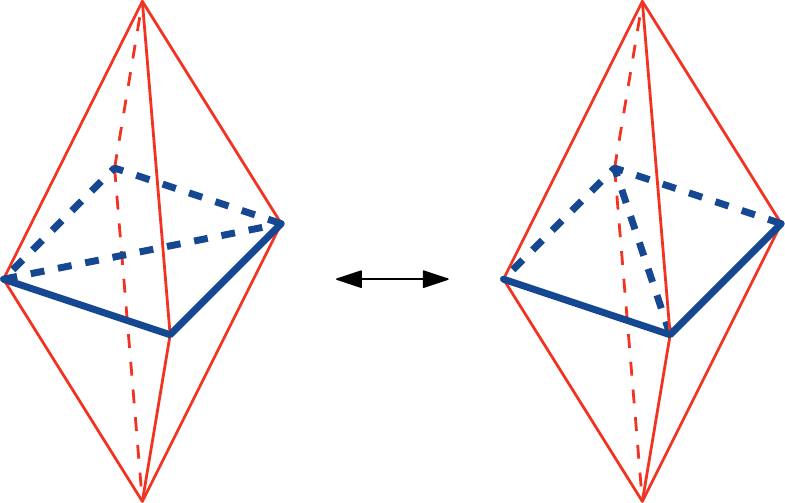}
    \\ \vspace{1em}
    \includegraphics[width=0.55\textwidth]{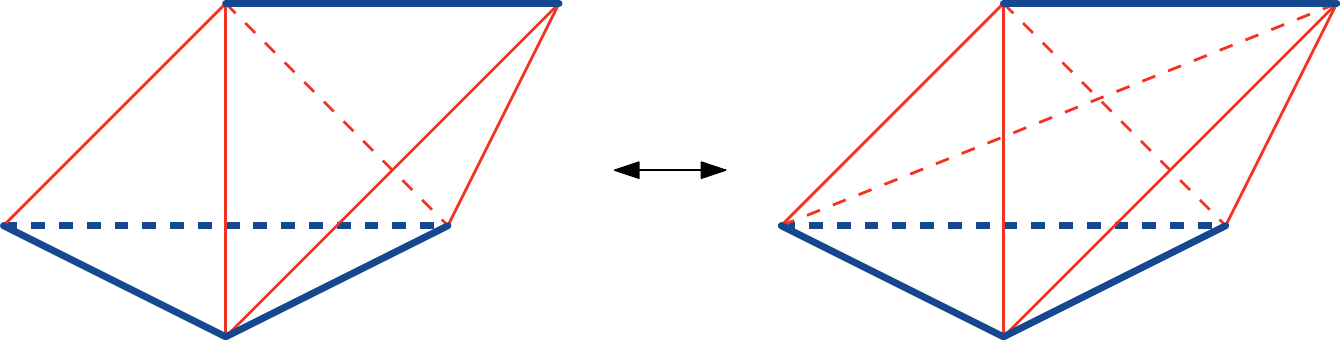}
    \caption{Three basic local moves in 3D CDT quantum gravity together with their inverses, and their effect on spatial slices. Space\-like edges are 
    drawn in blue and timelike ones in red. Top left: subdivision of a spatial triangle into three; top right: flip of a spatial edge. The move at the
    bottom does not affect the spatial triangulations (time inverse not shown).}
    \label{fig:moves-3d}
\end{figure}

Expectation values of geometric observables $\mathcal{O}$ in CDT quantum gravity are computed as 
\begin{equation}
	\langle \mathcal{O} \rangle = \frac{1}{Z}\, \sum_{T} \frac{1}{C_T}\,\mathcal{O}[T]\,  {\rm e}^{-S^\textrm{EH} [T]},
	\label{eq:q-exp-o}
\end{equation}
where $Z$ is the partition function defined in \eqref{cdtpi}. As already mentioned, the focus of our present work is a set of observables pertaining to
the two-dimensional spatial triangulations of constant integer time $t$ of the three-dimensional CDT configurations, which will be the subject of  
Sec.~\ref{sec:2d-qg}. Since it is not known how to compute $Z$ analytically in three dimensions, 
we will compute statistical estimates of the expectation value of an observable 
by sampling the CDT ensemble through Monte Carlo simulations.\footnote{Our implementation code can be found at \cite{brunekreef2022jorenb}.} In these simulations, we construct a random walk in the ensemble of CDT geometries by performing local updates (``moves'') on a triangulation. The basic set of 
moves we used is shown in Fig.\ \ref{fig:moves-3d}, see also \cite{ambjorn2001nonperturbative}. If we impose so-called detailed balance \cite{metropolis1953equation} on the updating procedure, by accepting or rejecting such moves with an appropriate probability, this random walk corresponds to a sample of the ensemble where geometries appear with a relative rate according to their Boltzmann weight. Since
subsequent geometries in a random walk are almost identical, we must perform a large number of local moves on a given geometry to obtain a new and
sufficiently independent one. This procedure is iterated to obtain a sequence of independent triangulations, and an estimate of the expectation value of 
an observable is computed as the weighted average \eqref{eq:q-exp-o} over this sequence. 
To study the continuum properties of observables, the number of building blocks should be taken to infinity. Since this is impossible in practice, due to the 
finiteness of our computational resources, we use finite-size scaling methods \cite{newman1999monte} to estimate the behaviour of the system in the continuum limit. For more details on computer simulations of three-dimensional CDT
we refer the interested reader to \cite{ambjorn2001computer}.

For a given value of the gravitational coupling $k_0$, the cosmological coupling $k_3$ is always tuned to its $k_0$-dependent pseudocritical 
value\footnote{which in the limit $N_3\!\rightarrow\!\infty$ would become the critical value $k_3^c(k_0)$}, which means that we are investigating a
one-dimensional phase space parametrized by $k_0$. The location of the critical point $k_0^c$ along this line, associated with the first-order
transition mentioned in Sec.\ \ref{sec:phases}, is not a universal quantity. For example, it depends on the regularity conditions imposed on the ensemble,
and the time extension $t_{\rm tot}$ of the geometries \cite{ambjorn2001nonperturbative}. 
In our analysis of the spatial slices we will use standard CDT simplicial manifolds\footnote{The effects of relaxing the local manifold constraints have been investigated further in \cite{brunekreef2022lorentzian}, where it was found that the order of the phase transition is likely unchanged, although the location of the critical point shifts to a smaller value as the restrictions are loosened.} and $t_{\rm tot}\! =\! 3$ with periodic boundary conditions in time, for which we 
have found $k_0^c\! \approx\! 6.24$.

\begin{figure}[t]
	\centering
	\includegraphics[width=0.5\textwidth]{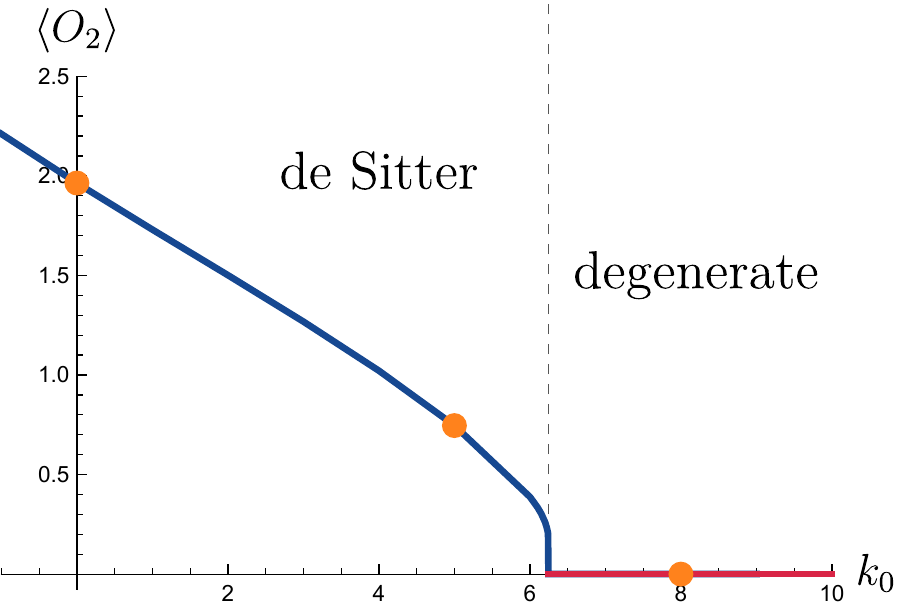}
	\caption{Expectation value of the order parameter $\optwo\! =\! N_{22}/N_{31}$ as a function of the bare coupling $k_0$, exhibiting a first-order phase transition 
	at $k_0^c\! \approx\! 6.24$. To the left of the transition is the de Sitter phase, and to its right the degenerate phase. Our measurements on
	spatial slices are taken for the $k_0$-values $0.0$, $5.0$ and $8.0$, as indicated.}
	\label{fig:phase-diagram}
\end{figure}

\begin{figure}[t]
	\centering
	\includegraphics[width=0.45\textwidth]{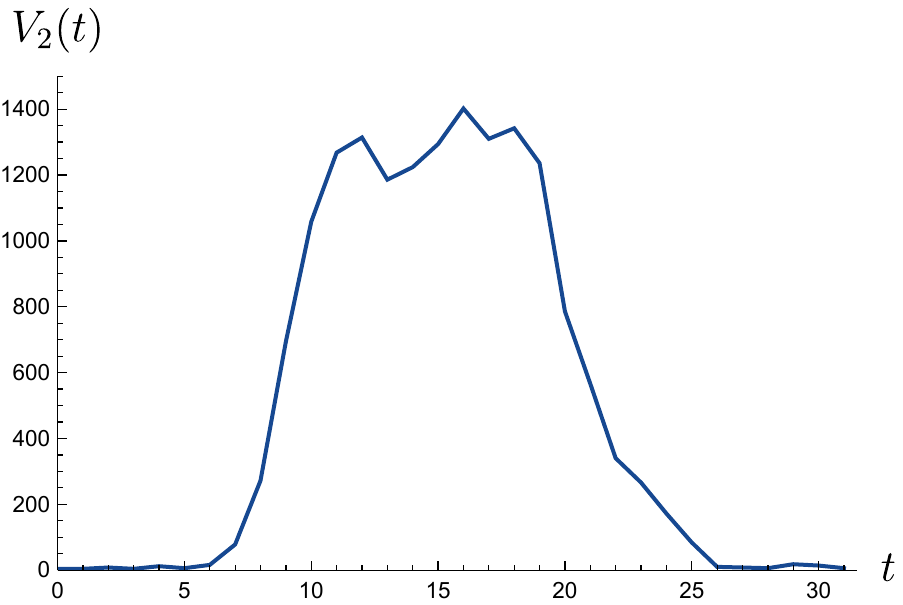}
	\hfill
	\includegraphics[width=0.45\textwidth]{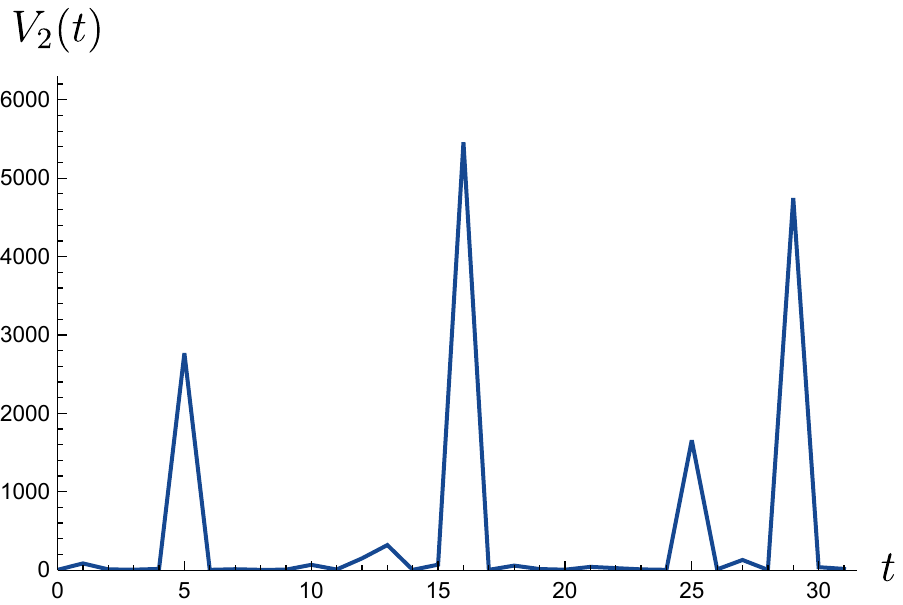}
	\caption{Volume profiles $V_2(t)$ of typical configurations appearing in the de Sitter phase (left) and the degenerate phase (right) of the
	three-dimensional CDT model on $S^1\!\times\! S^2$.}
	\label{fig:volprofs}
\end{figure}

A convenient order parameter to locate the phase transition is the ratio
\begin{equation}
	\optwo (T) = \frac{N_{22}(T)}{N_{31}(T)},
\end{equation}
where $N_{22}(T)$ and $N_{31}(T)$ denote the numbers of (2,2)- and (3,1)-simplices of the triangulation $T$ respectively. 
Its expectation value $\left\langle \optwo \right\rangle$ is
nonvanishing for small $k_0$ and drops to zero rapidly as the transition point $k_0^c$ is approached, beyond which it remains zero for all values 
of $k_0\! >\! k_0^c$ we have investigated. The measured values of $\left\langle \optwo \right\rangle$ as a function of $k_0$ are shown in 
Fig. \ref{fig:phase-diagram}, obtained in a system with $N_3\! =\! 64.000$ and $t_{\rm tot} = 3$.
The regions to the left and right of the transition correspond to the de Sitter and degenerate phases introduced earlier. Snapshots of typical
volume profiles $V_2(t)$, counting the number of triangles in the spatial slice at time $t$, are depicted in Fig.\ \ref{fig:volprofs} for a system with
$N_{31}\! =\! 16.000$ and $t_{\rm tot}\! =\! 32$.  
The volumes of neighbouring slices in the degenerate phase are largely uncorrelated, while they tend to align in the de Sitter phase.
When taking an ensemble average of the latter with the ``centres of volume'' aligned, the expectation value $\langle V_2(t)\rangle$ matches that of
a three-dimensional Euclidean de Sitter universe in a proper-time parametrization \cite{ambjorn2001nonperturbative}. 

We will perform measurements of the spatial slices for three different $k_0$-values, two in the de Sitter phase at $k_0\! =\! 0.0$ and $5.0$
and one in the degenerate phase at $k_0\! =\! 8.0$, cf.\ Fig.\ \ref{fig:phase-diagram}. The volumes $N_{31}$ of the systems we investigate take values in the range $[1.000, 96.000]$.
In choosing these particular values of $k_0$, we are staying away from the direct vicinity of the first-order transition at $k_0^c\! \approx\! 6.24$, to avoid
that the system jumps between the two phases during the simulation. The two points chosen in the de Sitter phase are spaced well apart, while taking into
account that the Monte Carlo algorithm becomes increasingly inefficient as $k_0$ is lowered. 
It may be worth pointing out that the expression for the discretized, bare Einstein-Hilbert action in \eqref{cdtpi} is highly non-unique, and that
the value $k_0\! =\! 0.0$ is therefore in no way physically distinguished.

The observed decoupling of neighbouring slices provides a strong argument for an effective slice
behaviour that is characteristic for the universality class of Euclidean dynamical triangulations, 
something our data in Sec.\ \ref{sec:2d-qg} below will confirm. 
As a cross check that the system's behaviour stays the same throughout the degenerate phase, we have performed a short series of 
measurements for all observables (except the curvature profile) at the much higher value of 
$k_0\! =\! 15.0$, which found no differences compared with $k_0\! =\! 8.0$.
Little is known about the slice behaviour in the de Sitter phase;
a preliminary investigation of the Hausdorff dimension for small slice volumes $V_2(t)\sim 1.000$
was made in \cite{ambjorn2001nonperturbative}, resulting in the estimate $d_H\! =\! 3.4\pm 0.4$. Furthermore,
a measure of homogeneity for the spatial slices was formulated and implemented in \cite{cooperman2014scaledependent}, but with inconclusive results. 

A final technical issue to be discussed before turning to a description of the measurement results is that of volume-fixing. 
As usual, we perform all measurements at fixed spacetime volume; more precisely, since the Monte Carlo moves are not volume-preserving, 
the simulations are run in the vicinity of a target volume. The latter can be stated in terms of the total volume $N_3$ or 
in terms of $N_{31}$, which is what we will do below. Both prescriptions are essentially equivalent, since in the case of
periodic boundary conditions in time we have the identity $N_3\! =\! 2 N_{31} + N_{22}$, and since for fixed $k_0$ the ratio between
$N_{22}$ and $N_{31}$ is approximately constant, and can be read off the graph in Fig.\ \ref{fig:phase-diagram}. Note also that
$N_{31}$ is equal to the total number of \emph{spatial} triangles in the triangulation, i.e.\ the sum over all $t$ of $V_2(t)$.
The approximate volume-fixing is implemented by adding a term
\begin{equation}
	S_\textrm{fix}[T] = \epsilon\, \big(N_{31}(T)-\tilde{N}_{31}\big)^2
\label{vcon}	
\end{equation}
to the bare action, where $\tilde{N}_{31}$ denotes the desired target volume and the value of the small, positive parameter $\epsilon$ determines the
typical size of the fluctuations of $N_{31}$ around $\tilde{N}_{31}$. In the simulations performed for this work, we generally set $\epsilon$ to values on the order of $10^{-5}$.

In addition, since we are interested in the intrinsic properties of the spatial slices and in extracting their continuum behaviour from finite-size scaling,
we must collect measurement data at different, fixed \emph{slice} volumes.
Instead of adding further volume-fixing terms for the individual slices to the action, which would run the risk of introducing an unwanted bias, 
we let the individual slices fluctuate freely, subject only to the total volume constraint (\ref{vcon}), but take data only when a slice hits
a precise desired value $\tilde{V}_2$.
More concretely, in between two measurements we first perform a fixed number of attempted moves\footnote{These attempted moves will be accepted or rejected according to the detailed balance condition mentioned earlier.}, collectively referred to as a \emph{sweep}. Different observables can have different autocorrelation times (measured in Monte Carlo steps) and therefore require different sweep sizes. A typical sweep size is taken to be on the order of 1.000 times the target 
volume $\tilde{N}_{31}$. After completion of a sweep, we continue performing local updates until one of the slices hits the target two-volume $\tilde{V}_2$. 
We then perform a measurement of the observable under consideration on this slice, and subsequently start a new sweep.

The choice of the target three-volume $\tilde{N}_{31}$ that maximizes the probability of encountering slices with target two-volume $\tilde{V}_2$
depends on the phase. In the de Sitter phase and for the small time extension $t_{\rm tot}$ we have used, the total volume spreads roughly evenly 
over the available slices and an appropriate choice is $\tilde{N}_{31} \! = \! t_{\rm tot} \cdot \tilde{V}_2$. In the degenerate phase, the volume tends to
concentrate on one of the slices, and a good choice is $\tilde{N}_{31}\! =\! \tilde{N}_2 + m$, where for $t_{\rm tot}\! =\! 3$ and $\tilde{V}_2\! >\! 1.000$
we found that $m\! =\! 100$ is a convenient choice.

To maximize the volume of the spatial slices, we work with $t_{\rm tot}\! =\! 3$, 
the minimal number allowed by our simulation code. 
Both this choice and our choice of periodic boundary in time can in principle have an influence on the behaviour of observables, even if our
measurements are confined to individual slices. Investigations of the transfer matrix in four-dimensional CDT quantum gravity 
have indicated that such a set-up can be appropriate, at least for selected observables \cite{ambjorn2013transfer}.
As an extra check, we have performed a few measurements on systems with larger $t_{\rm tot}\!\lesssim\! 32$, and found 
the same behaviour for the slice geometries. This is reassuring, but not a substitute for a more systematic investigation of the influence of
these global choices, which goes beyond the scope of our present work. This proviso should be kept in mind when interpreting the outcomes 
of our research, which will be presented next.

\section{Geometric observables on spatial hypersurfaces}
\label{sec:2d-qg}

The main objective of our work is a detailed measurement of the geometric properties of the two-dimensional spatial hypersurfaces
of constant integer proper time $t$ in three-dimensional CDT quantum gravity, both in the de Sitter and the degenerate phase. 
We will compare our findings with known results for nonperturbative
models of two-dimensional quantum gravity. There are two pure-gravity systems in $D\! =\! 2$ available for reference, Euclidean DT and Lorentzian CDT
quantum gravity. However, especially in the de Sitter phase there are no stringent reasons why the slice geometries should match these known
systems, since they are part of a larger, three-dimensional geometry. For example, the ambient geometry may induce extrinsic curvature terms 
on the spatial slices, which are not present in intrinsically two-dimensional situations. 
The following subsections will deal in turn with the four quantities we have studied on the spatial slices: the vertex order, the entropy exponent, 
the Hausdorff dimension and the curvature profile.

\subsection{Vertex order}
\label{subsec:coord}

We first examined the distribution of the vertex order, which counts the number $q(v)$ of spacelike edges meeting at a given vertex $v$.
For the simplicial manifolds we use in the simulations, the possible vertex orders are $q(v)\! =\! 3,4,5,\dots $.
As already mentioned earlier, the distribution of the $q(v)$ is not strictly speaking an observable. It is a non-universal property of the discrete lattice, which  
depends on the details of the lattice discretization and does not have an obvious continuum counterpart. For example, using  
quadrangulations instead of triangulations leads to the same continuum theory of two-dimensional Euclidean DT quantum gravity \cite{miermont2013brownian}, 
but the two models have different distributions of vertex orders. We have studied this quantity nevertheless, since it is known analytically for both the DT and CDT ensembles and gives us a first idea of whether and how our system changes as a function of the bare coupling $k_0$. 

\begin{figure}[t]
	\centering
	\includegraphics[width=0.6\textwidth]{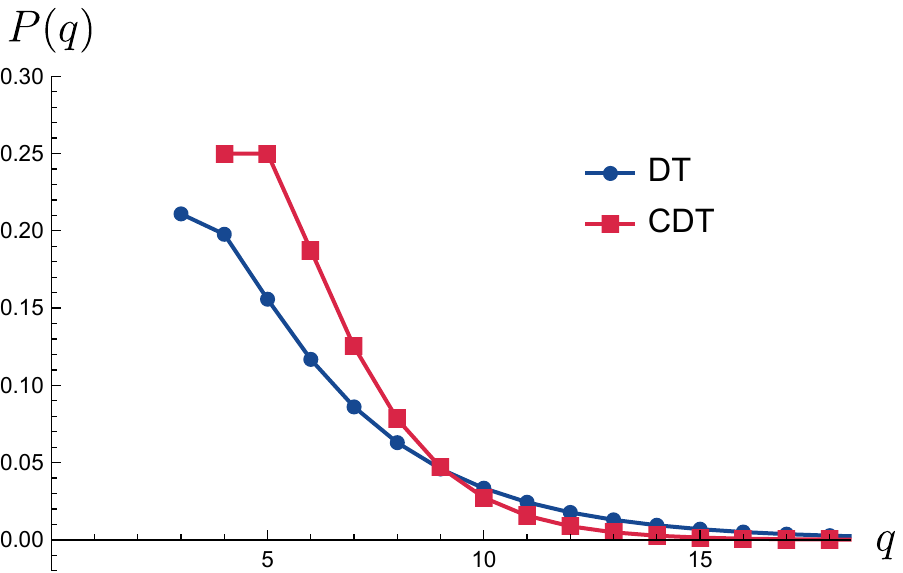}
	\caption{Probability distribution of the vertex order $q$ for the ensembles of two-dimensional Euclidean DT and Lorentzian CDT quantum gravity, in the
	limit of infinitely many triangles.}
	\label{fig:coord-theory}
\end{figure}

The normalized probability distribution $P(q)$ for the vertex order in two-dimensional DT quantum gravity in the thermodynamic limit
has been determined analytically as \cite{boulatov1986analytical}
\begin{equation}
	P_\textrm{DT}(q) = 16 \left(\frac{3}{16}\right)^q  \frac{(q-2)(2q-2)!}{q!(q-1)!}, \quad \quad q \geq 2,
	\label{eq:dt-coord}
\end{equation}
with a large-$q$ behaviour $\sim\! \left(\frac{3}{4}\right)^q$. The corresponding probability distribution for two-dimen\-sio\-nal CDT quantum gravity is given by \cite{ambjorn1999new}
\begin{equation}
	P_\textrm{CDT}(q) = \frac{q-3}{2^{q-2}}, \quad \quad q \geq 4,
    \label{eq:cdt-coord}
\end{equation}
with a fall-off behaviour $\sim\! \left(\frac{1}{2}\right)^q$ for large $q$. Both distributions are shown in Fig.\ \ref{fig:coord-theory}.

\begin{figure}[t]
	\centering
	\includegraphics[width=0.6\textwidth]{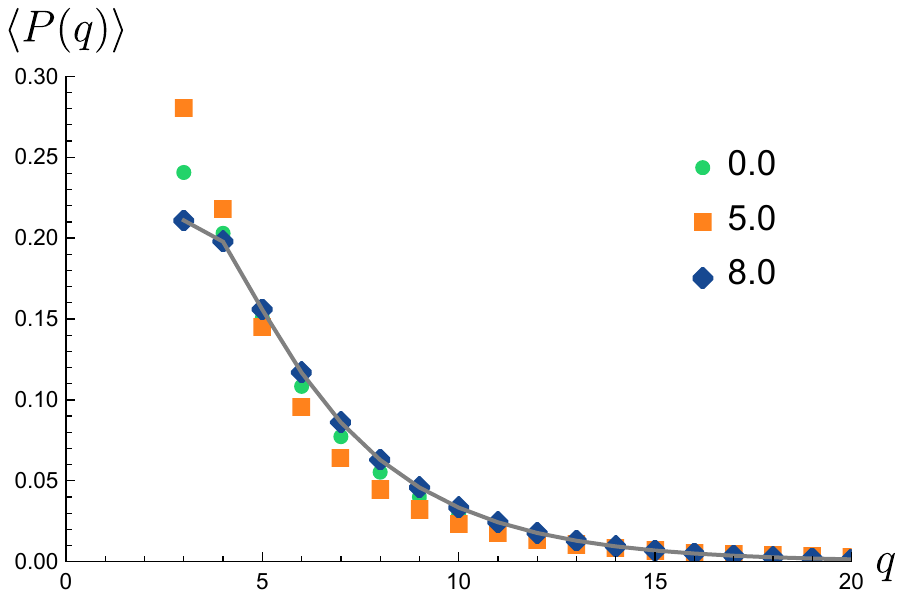}
	\caption{Measured probability distribution of the vertex order $q$ on spatial slices of volume $V_2\! =\! 4.000$ in three-dimensional CDT, for 
	$k_0\! =\! 0.0$, $5.0$ and $8.0$
	and $q\! \leq\! 20$. For comparison, the line connecting the exact values for DT from Fig.\ \ref{fig:coord-theory} are included.}
	\label{fig:cdt-coord-results-nonlog}
\end{figure}

\begin{figure}[t]
	\centering
	\includegraphics[width=0.6\textwidth]{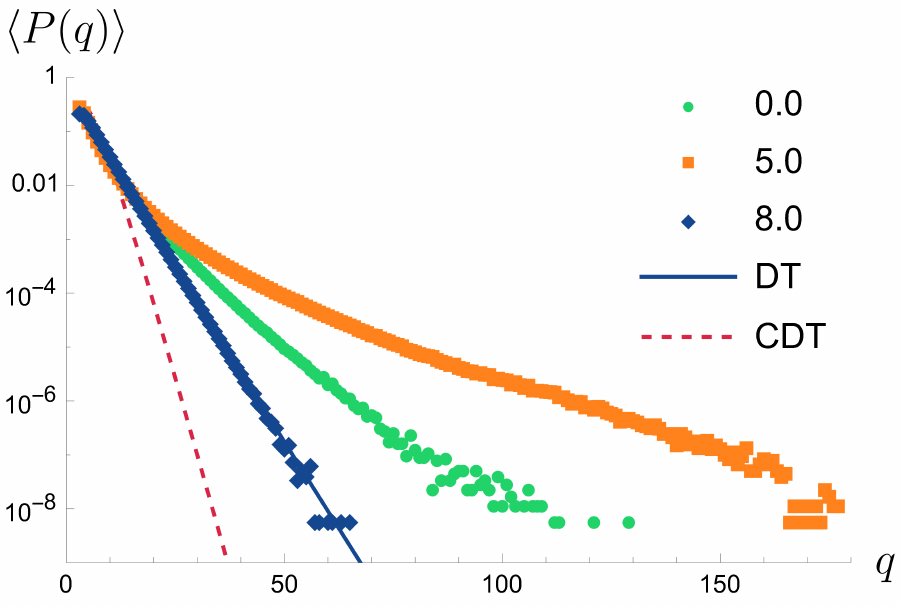}
	\caption{Measured probability distribution of the vertex order $q$ (logarithmic scale) on spatial slices of volume $V_2\! =\! 4.000$ in three-dimensional CDT, for $k_0\! =\! 0.0$, $5.0$ and $8.0$. The unbroken straight line is the analytical result \eqref{eq:dt-coord} for two-dimensional DT, and the dashed straight line the analytical result \eqref{eq:cdt-coord} for two-dimensional CDT.}
	\label{fig:cdt-coord-results}
\end{figure}

We took measurements at the three chosen phase space points $k_0 = 0.0, 5.0$ and $8.0$, with a target volume $\tilde{V}_2\! =\! 4.000$ for the spatial slices,
corresponding to a target three-volume $\targn\! =\! 12.000$ in the de Sitter phase, and $\targn\! =\! 4.400$ in the degenerate phase. 
The sweep size was set to $100 \cdot \targn$ in each case. For each value of $k_0$, we collected measurements of $P(q)$ for 100k different slices, 
by recording for each slice the full set of vertex orders $q(v)$ for all vertices and normalizing the resulting histogram. We then approximated the
eigenvalue $\langle P(q)\rangle$ by taking the ensemble average over this data set according to the prescription (\ref{eq:q-exp-o}). 
The results of the measurements for $q\!\in\! [3,20]$ are shown in Fig.\ \ref{fig:cdt-coord-results-nonlog}, and are clearly distinct for the three
$k_0$-values. While the distribution for $k_0\! =\! 8.0$ in the degenerate phase is a very good match for the analytical result, 
the distributions for $k_0\! =\! 0.0$ and $5.0$ in the de Sitter phase are not good matches and are also different from each other.
This is confirmed when plotting the measurements on a logarithmic scale, taking into account a much larger range of $q\!\leq \! 180$,
as depicted in Fig.\ \ref{fig:cdt-coord-results}, which also includes the known 
distributions for Euclidean DT and CDT as thin straight lines.
We see that within measurement accuracy, the distribution of vertex orders in the degenerate phase of the model is indistinguishable from 
\eqref{eq:dt-coord}, a result that is also compatible with the numerical simulations of pure Euclidean DT quantum gravity 
performed in \cite{boulatov1986analytical}. By contrast, the distributions obtained in the de Sitter phase exhibit a very different behaviour, at least for large $q$.
High-order vertices are relatively speaking more probable, and the data cannot be fitted to a single exponential over the $q$-range we have explored. 
Moreover, unlike what happens in the degenerate phase, the distributions within the de Sitter phase depend on the value of $k_0$.

To obtain a more detailed picture of the dependence on $k_0$, 
we performed a series of shorter simulation runs at additional points in the de Sitter phase. 
When entering the de Sitter phase from the degenerate phase by crossing the phase transition at $k_0^c$, the vertex order distribution 
jumps discontinuously to a shape similar to that for $k_0\! =\! 5.0$. 
As $k_0$ is decreased further inside this phase, the distribution changes shape in a continuous way; the distributions we measured for points 
in the interval $0.0\! < \! k_0\! <\! 5.0$ interpolate in a straightforward manner between the ones for $k_0 = 0.0$ and $k_0 = 5.0$ 
shown in Figs.\ \ref{fig:cdt-coord-results-nonlog} and \ref{fig:cdt-coord-results}. 

To summarize, the measurements of the vertex order distribution in the degenerate phase produce an excellent match with the known one for
DT quantum gravity. By contrast, the distributions found in the de Sitter phase do not match those of the standard DT or CDT ensembles in $D\! =\! 2$.
As mentioned earlier, this does not necessarily mean that the slice geometries in the de Sitter phase do not lie in either of the associated universality classes, 
but it is a first indication that they may not. By looking at genuine observables next, we will be able to make more definite statements about the
universal geometric properties of the slice geometries.

\subsection{Entropy exponent}
\label{entresults}

An important parameter characterizing two-dimensional systems of random geometry is the entropy exponent $\gamma$,
which contains information about the behaviour of the partition function at fixed two-volume $N_2$ (number of triangles), in the limit as $N_2$ 
becomes large.\footnote{We use $N_2$ to denote two-volumes in two-dimensional models of quantum gravity and $V_2$ to denote
the two-volume of a spatial slice in three-dimensional quantum gravity.}
Recall that the path integral of DT quantum gravity in $D\! =\! 2$ with bare cosmological constant $\lambda$ can be written as the infinite sum
\begin{equation}
	Z(\lambda ) = \sum_{N_2} Z(N_2)\, {\rm e}^{-\lambda N_2},
\label{z2part}
\end{equation}
which is the (discrete) Laplace transform of the partition function $Z(N_2)$ for fixed volume. For large $N_2$, $Z(N_2)$ behaves like
\begin{equation}
	Z(N_2) \sim {\rm e}^{\lambda^c N_2} N_2^{\gamma-3} \left(1+{\mathcal O}(1/N_2)\right),
	\label{eq:fixed-n-partition}
\end{equation}
whose leading exponential growth is governed by a (non-universal) critical cosmological constant $\lambda^c\! >\! 0$ and whose
subleading power-law behaviour defines the universal entropy exponent $\gamma$, which for DT quantum gravity is given by $\gamma\! =\! -1/2$.
The asymptotic functional form (\ref{eq:fixed-n-partition}) continues to hold when conformal matter of central charge $c\! <\! 1$ is added to the Euclidean
quantum gravity model, giving rise to the entropy exponents \cite{knizhnik1988fractal}
\begin{equation}
	\gamma = \frac{1}{12} \left(c-1 - \sqrt{(25-c)(1-c)}\right), 
	\label{eq:gamma_s}
\end{equation}
with $c\! =\! 0$ corresponding to the pure-gravity case. Two-dimensional CDT quantum gravity, which is not described by formula (\ref{eq:gamma_s}), 
is characterized by $\gamma\! = \! 1/2$
\cite{ambjorn1998nonperturbative}. 
Computer simulations have demonstrated that adding matter with $c\! =\! 4$ to the CDT system induces a phase transition in the 
geometry \cite{ambjorn2000crossing}, but the corresponding entropy exponent is not known.

According to \cite{jain1992worldsheet}, the distribution of so-called baby universes 
in two-dimensional Euclidean quantum gravity -- parts of a geometry that are connected to the larger bulk geometry via a thin neck -- 
depends in a simple way on the 
entropy exponent $\gamma$. This insight was used subsequently to formulate a prescription of how to extract $\gamma$ by measuring the
distribution of minimal-neck baby universes (``minbus'') in the DT ensemble with the help of Monte Carlo simulations \cite{ambjorn1993baby}. 
A minbu is a simply connected subset of disk topology
of a two-dimensional triangulation $T$, which is connected to the rest of $T$ along a loop consisting of three edges, 
which is the minimal circumference of  a neck allowed by the simplicial manifold conditions.  

As shown in \cite{jain1992worldsheet,ambjorn1993baby}, it follows from relation (\ref{eq:fixed-n-partition}) that the average number 
$\bar{n}_{N_2}(B)$ of minbus of volume $B$ (counting the number of triangles in the minbu) in a spherical triangulation of 
volume $N_2$ for sufficiently large $B$, $N_2$ behaves like 
\begin{equation}
	\label{eq:minbu-dist}
	\bar{n}_{N_2} (B) \sim (N_2-B)^{\gamma-2} B^{\gamma-2}.
\end{equation}
By measuring the distribution of minbus across a range of volumes $B$ for fixed $N_2$ in a DT ensemble and fitting the results to the function
(\ref{eq:minbu-dist}), the expected results $\gamma\! =\! -1/2$ for pure gravity and $\gamma\! =\! -1/3$ for gravity coupled to Ising spins ($c\! =\! 1/2$)
were reproduced within measuring accuracy \cite{ambjorn1993baby}. 

We have carried out a similar analysis on the spatial slices of triangulations generated by Monte Carlo simulations of 
three-dimensional CDT quantum gravity. There is no obvious reason why (\ref{eq:fixed-n-partition}) should hold for some ``effective'' fixed-volume
partition function for the two-volume $N_2\! =\! V_2$ of a single spatial slice in this three-dimensional system. However, if the number of (2,2)-simplices
drops essentially to zero \emph{and} neighbouring slices decouple, as is the case in the degenerate phase, the three-dimensional partition function 
at fixed volume will depend only on $N_{31}$ (equal to the total two-volume), which makes it plausible that (\ref{eq:fixed-n-partition}) holds on 
individual spatial slices, with $N_2\! =\! V_2$. In the de Sitter phase, if the spatial geometries can be described in terms of two-dimensional DT 
quantum gravity, the minbu method will presumably also lead to $\gamma\! =\! -1/2$.

\begin{figure}[t]
	\centering
	\includegraphics[width=0.45\textwidth]{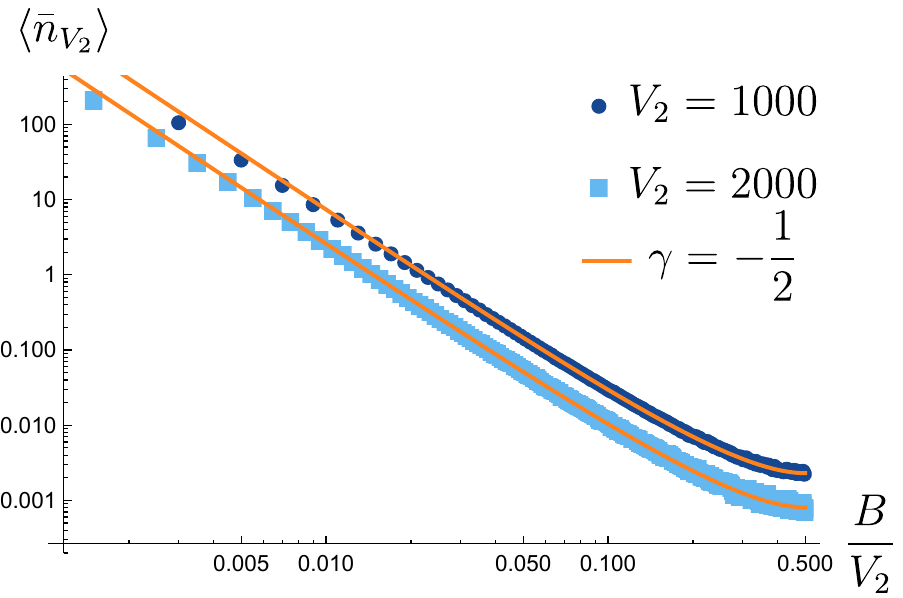}
	\\ \vspace{1cm}
	\includegraphics[width=0.45\textwidth]{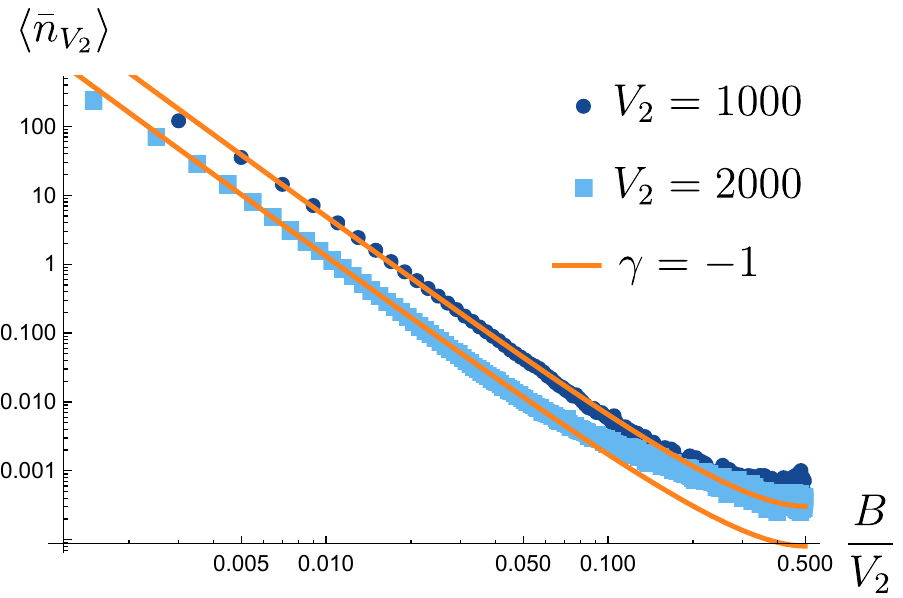}
	\hfill
	\includegraphics[width=0.45\textwidth]{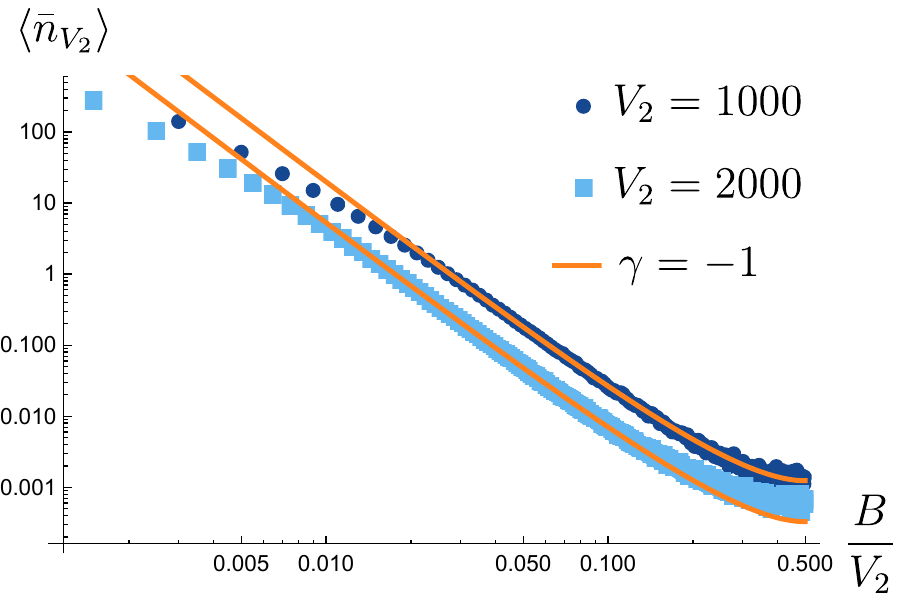}
	\caption{Expectation values of the distribution $\bar{n}_{V_2}$ of minbu sizes for spatial slices of volume $V_2\! =\! 1.000$ and $2.000$
	in three-dimensional CDT configurations, in the degenerate phase ($k_0\! =\! 8.0$, top) and the de Sitter phase ($k_0\! =\! 0.0$, bottom left,
	and $k_0\! =\! 5.0$, bottom right), on a log-log scale. The continuous lines are best fits of the form \eqref{eq:minbu-dist} for specific values of 
	$\gamma$.
	Error bars are smaller than the dot size.}
	\label{fig:minbu-dist}
\end{figure}

We measured the distribution $ \bar{n}_{V_2}(B)$ of minbu sizes $B$ for target slice volumes $\tilde{V}_2\! =\! 1.000$ and $2.000$ 
at the three phase space points $k_0 = 0.0, 5.0$ and $8.0$.
The sweep size was set to $10^4\! \cdot\! \tilde{V}_2$ for measurements in the degenerate phase, and $10^5\! \cdot\! \tilde{V}_2$ in the de Sitter phase. 
We used longer sweeps in the de Sitter phase because the observed autocorrelations were much larger, especially at $k_0\! =\! 0.0$, 
where the algorithm is much less efficient. We collected on the order of $5\! \cdot\! 10^4$ minbu size histograms for 
$\tilde{V}_2\! =\! 1.000$ and $1.5\! \cdot\! 10^5$ histograms for $\tilde{V}_2\! =\! 2.000$ in the de Sitter phase, and
$4 \cdot 10^5$ histograms for $\tilde{V}_2\! =\! 1.000$ and $2 \cdot 10^5$ histograms for $\tilde{V}_2\! =\! 2.000$ in the degenerate phase.
The resulting expectation values $\langle \bar{n}_{V_2}\rangle$ of the minbu size distribution 
as a function of the normalized ratio $B/V_2\in [0,1/2]$ are shown in Fig.\ \ref{fig:minbu-dist}, together with
best fits of the form \eqref{eq:minbu-dist} for specific values of $\gamma$. The best fits were determined following the procedure used 
in \cite{ambjorn1993baby}, and involved a subleading correction term to the power law $B^{\gamma -2}$, as is described in more detail in
Appendix \ref{app:entropy} below.

In the degenerate phase (Fig.\ \ref{fig:minbu-dist}, top), the choice $\gamma\! =\! -1/2$ fits the data extremely well throughout the entire range
of $B/V_2$, with the exception of the smallest minbu sizes. Our results coincide within error bars with those reported in Table 1 
of \cite{ambjorn1993baby}. This confirms that the spatial slices in the degenerate phase exhibit behaviour consistent with DT quantum gravity in two
dimensions. 

In the de Sitter phase, there is no $\gamma$-value that leads to a good fit over the full range of $B/V_2$,
even when we disregard the region of small minbu size $B$, where \eqref{eq:minbu-dist} is known to be inaccurate. 
The plots for $k\! =\! 0.0$ and $k_0\! =\! 5.0$ (Fig.\ \ref{fig:minbu-dist}, bottom) illustrate the optimum of what can be achieved, namely,
a fit that works reasonably well for an intermediate range of $B/V_2$, in this case, a fit corresponding to $\gamma\! =\! -1$.
However, this clearly does not fit the data for large minbus near $B/V_2\!=\! 1/2$, especially not for the smaller value of $k_0$, 
and the discrepancy seems to get worse with increasing volume $V_2$. 

We conclude that the minbu distribution in the de Sitter phase does not follow the functional form of the right-hand side of relation \eqref{eq:minbu-dist},
at least not for the slice volumes we have considered (and which seem to be sufficient in the degenerate phase). 
A possible explanation is that the ``effective'' partition function $Z(V_2)$, which is obtained from the three-dimensional CDT partition function $Z(N_3)$ for
fixed three-volume by integrating out all degrees of freedom except for a single slice volume $V_2$, does not have the asymptotic form (\ref{eq:fixed-n-partition}).
This would imply that it is not in the universality class of a DT model with central charge $c\! <\! 1$ or of CDT quantum gravity. 
A more subtle scenario would be that
$Z(V_2)$ does behave according to \eqref{eq:fixed-n-partition} (and perhaps does correspond to a known gravity model in $D\! =\! 2$), but that the 
derivation of the minbu distribution \eqref{eq:minbu-dist} is invalidated by the presence of correlations between the bulk and a baby universe 
that exist because of the embedding of the spatial slice in a three-dimensional simplicial geometry. Such correlations are not present in the
degenerate phase because of the absence of (2,2)-simplices. In the following two subsections, we will look at observables which also characterize the
intrinsic geometry of the spatial slices, but whose determination is less subtle than that of the entropy exponent.

\subsection{Hausdorff dimension}

The Hausdorff dimension is a notion of fractal dimension that can be used to characterize a quantum geometry in an invariant manner. Broadly speaking,
it is extracted by comparing volumes with their characteristic linear size, measured in terms of a geodesic distance. There have been many studies of
the Hausdorff dimension in the context of DT and CDT quantum gravity, including extensive investigations 
in two-dimensional Euclidean DT models with and without matter
(see, for example, \cite{ambjorn1995scaling,ambjorn1995fractal,ambjorn1997quantumb,barkley2019precision} and references therein). Following \cite{ambjorn1995scaling}, we will investigate a local and a global (``cosmological'') variant of this observable on the spatial slices. 
From analytical considerations, it is known that in two-dimensional Euclidean DT quantum gravity without matter 
both types of Hausdorff dimension are equal to four
(i.e.\ different from the topological dimension of the triangular building blocks),
while for two-dimensional CDT quantum gravity they are both equal to two \cite{ambjorn1998nonperturbative,durhuus2010spectral}.

When measuring the Hausdorff dimension numerically, one can use either the link distance or the dual link distance as a discrete implementation
of the geodesic distance. In known systems of pure gravity in $D\! =\! 2$, they lead to equivalent notions of geodesic distance in the continuum limit, but
for finite lattice sizes, one particular choice may be more convenient. This is true for our investigation below, where we will use the dual link distance, which is defined
between dual vertices (equivalently, centres of triangles) and given by the length of the shortest path along dual links between the vertices. 

When comparing our results with previous measurements of the Hausdorff dimension in the context of two-dimensional Euclidean DT \cite{catterall1995scaling,ambjorn1995fractal}, one must take into account that the latter employed a larger ensemble of geometries.
This generalized ensemble allows for local gluings of the equilateral triangles that violate the strict simplicial manifold conditions\footnote{two triangles
cannot share more than one edge, any two vertices cannot be connected by more than one edge}. When characterizing the triangulations in terms of their dual, trivalent graphs, 
the generalization consists in allowing for tadpole and self-energy insertions. It has been demonstrated to reduce finite-size effects \cite{ambjorn1995new},
and is justified by the fact that the model on the enlarged ensemble can be shown to lie in the same universality class (see e.g.\ \cite{schneider1999universality}
and references therein). Since no analogous result is available in three-dimensional CDT quantum gravity, it is prudent to use only simplicial manifolds,
which implies that the spatial slices are simplicial manifolds too. This may affect the quality of our results, compared with the earlier,
purely two-dimensional investigations.
Note also that nonlocal minbu surgery moves were used in \cite{ambjorn1995fractal} to complement the standard local Monte Carlo moves and reduce
autocorrelation times, something we cannot easily implement on two-dimensional embedded triangulations.

\subsubsection{Local Hausdorff dimension}
\label{sssec:mhd}

A key quantity in determining the local Hausdorff dimension $d_h$ of a two-dimensional triangulation $T$ is the shell volume $S(r)$,
which in our implementation counts the number of dual vertices (equivalently, triangles) at dual link distance $r$ from a given dual vertex.   
The corresponding observable is the quantity $\bar{S}(r)$, obtained by averaging $S(r)$ 
over all dual vertices of $T$. 
The reason for using the dual link distance is that shells with respect to the link distance quickly cover a large fraction of the geometry as $r$ grows.
This implies that the average shell volumes $\bar{S}(r)$ cover only a small range of radii $r$ before dropping to zero,
yielding too few data points to make reliable estimates of either Hausdorff dimension.
The local Hausdorff dimension is extracted from
the expectation value $\bar{S}(r)$ in the ensemble at fixed two-volume $N_2$ according to 
\begin{equation}
\langle \bar{S}(r)\rangle_{N_2} \sim r^{d_h -1},
\label{eq:haus-micro}
\end{equation} 
for small $r$. In other words, $d_h$ captures the initial, volume-independent power-law growth of small geodesic spherical shells, 
where $r$ must be sufficiently large
to avoid dominance by discretization artefacts and sufficiently small to avoid significant corrections to the simple power law behaviour (\ref{eq:haus-micro}),
if such a behaviour is indeed present.

We have measured the expectation values of average shell volumes as a function of the dual link distance $r$, 
at slice volumes $V_2\! =\! 16k$ and $32k$ and
at the three chosen phase space points $k_0$, which is all straightforward. The local Hausdorff dimension $d_h$ was extracted
by fitting the measured data to the functional form
\begin{equation}
\langle \bar{S}(r)\rangle_{V_2}	 = c \cdot (r+a)^{d_h-1},
	\label{eq:haus-micro-fit}
\end{equation}
where -- following \cite{ambjorn1995fractal} -- we have introduced an offset $a$ to account for short-distance discretization artefacts,  and
$c$ is a multiplicative parameter.
The dependence of the Hausdorff dimension on the chosen fitting range $r\!\in\! [r_{\textrm{min}},r_{\textrm{max}}]$ will be analyzed in more detail below.

\begin{figure}[t]
	\centering
	\includegraphics[width=0.45\textwidth]{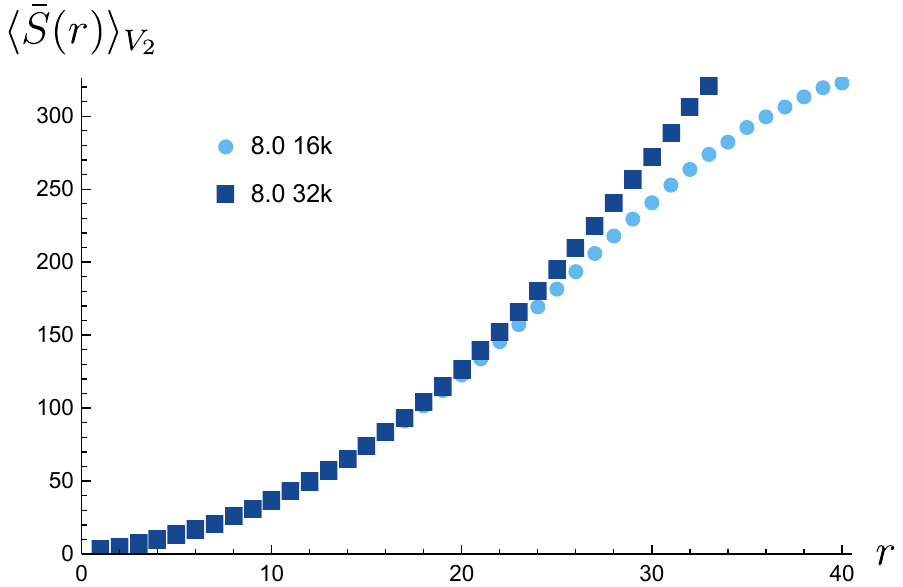}
	\hfill
	\includegraphics[width=0.45\textwidth]{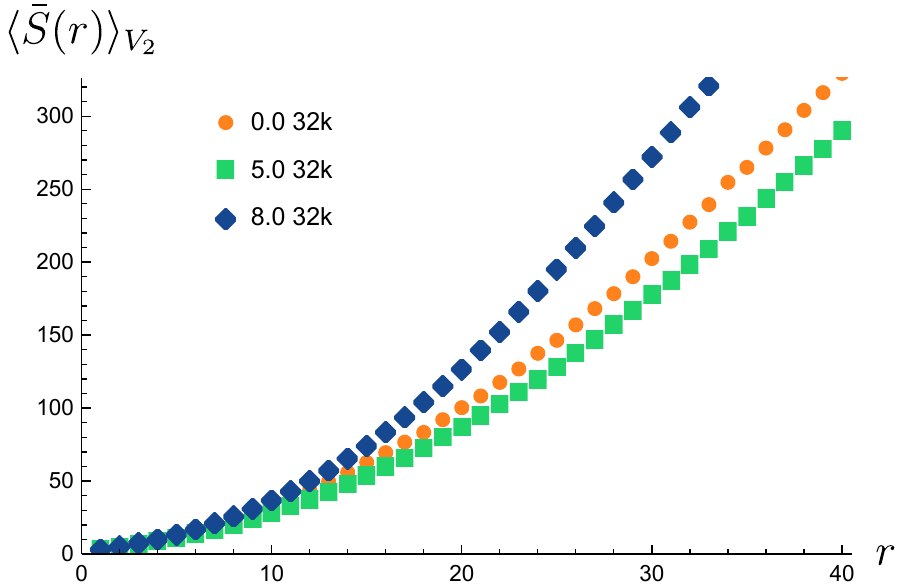}
	\caption{Expectation values $\langle \bar{S}(r)\rangle_{V_2}$ of average shell volumes in the degenerate phase $(k_0\! =\! 8.0)$ for slice volumes $V_2\! = \!16k, 32k$ (left), and for all three phase space points $k_0\! =\! 0.0$, $5.0$ and $8.0$ for slice volume $V_2\! =\! 32k$ (right). Error bars are smaller than
	dot size.}
	\label{fig:haus-micro-all}
\end{figure}

The measured expectation values of the average shell volume are shown in Fig.\ \ref{fig:haus-micro-all} for $0\! \leq\! r\! \leq\! 40$. 
The plot on the left, describing the behaviour of the system in the degenerate phase, illustrates the difference between the data for slice volumes
$16k$ and $32k$. The data points for the smaller volume start deviating from the common behaviour around $r\!\approx\! 20$, 
indicating that a power-law fit becomes
inadequate beyond this point. The plot on the right illustrates the dependence of the initial slope on the value of $k_0$. Note that although the curve for
$k_0\! =\! 0.0$ lies in between the two other curves, the corresponding Hausdorff dimension obtained from fitting to the functional form
\eqref{eq:haus-micro-fit} comes out lower than that for $k_0\! =\! 5.0$, see below. 

Since the criteria for fixing the fitting range $r\!\in\! [r_{\textrm{min}},r_{\textrm{max}}]$ for eq.\ \eqref{eq:haus-micro-fit} are only approximate, it is important to understand which choice is most appropriate and how stable the results for $d_h$ are when the range is varied. 
Some earlier work used the range $r\!\in\! [5,15]$ in terms of the dual link distance, and for a system of volume $N_2\! =\! 64k$
\cite{ambjorn1998quantum}.

\begin{figure}[t]
	\centering
	\includegraphics[width=0.45\textwidth]{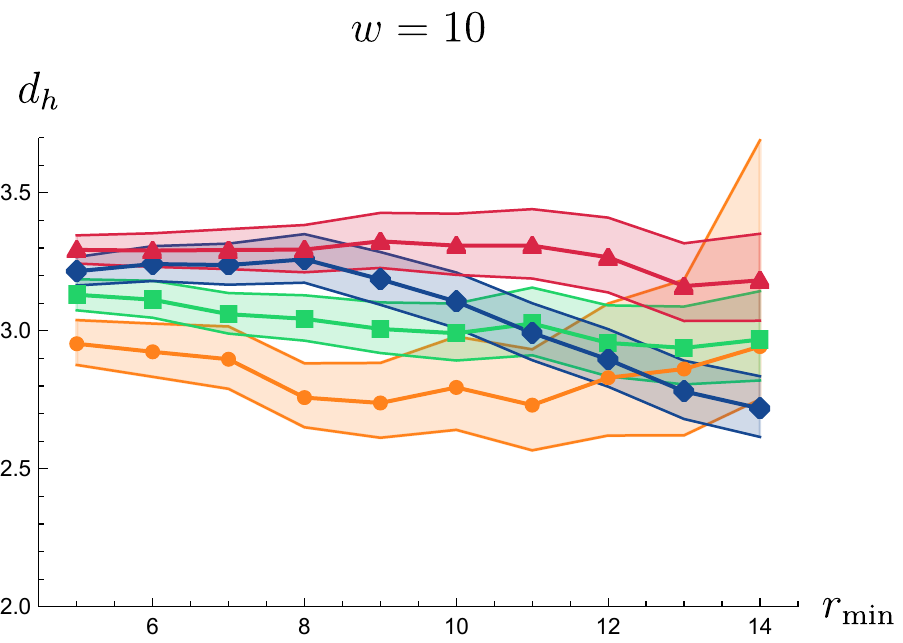}
	\hfill
	\includegraphics[width=0.45\textwidth]{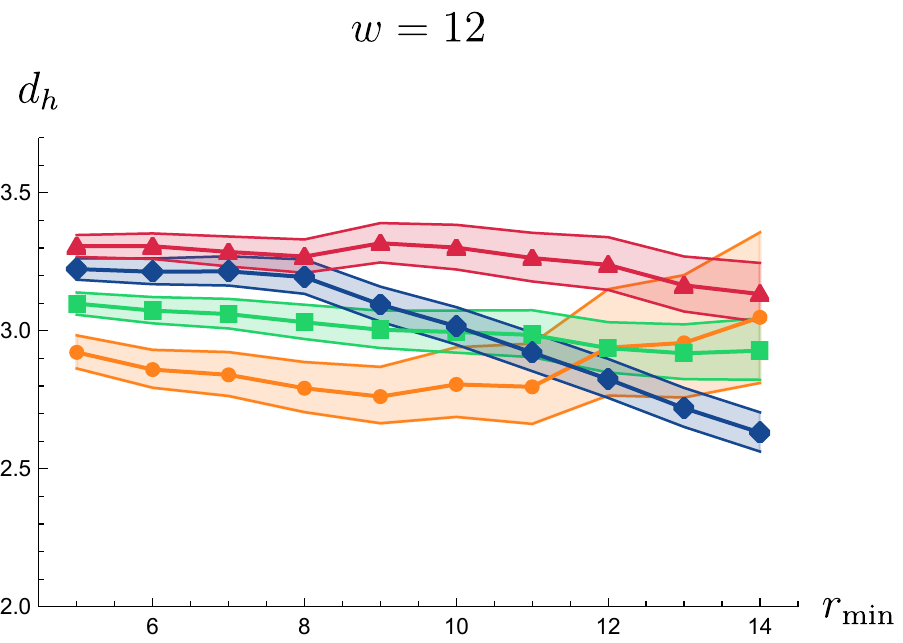}
	\\ \vspace{1em}
	\includegraphics[width=0.4\textwidth]{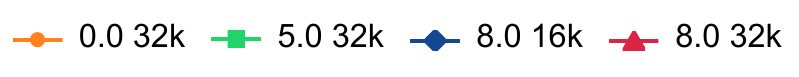}
	\caption{Best fit values for $d_h$ from fitting \eqref{eq:haus-micro-fit} to the measured expectation values $\langle \bar{S}(r)\rangle_{V_2}$ in the range $r\!\in\! [r_\textrm{min}, r_\textrm{min}\! +\! w]$ for spatial slices in the degenerate phase ($k_0\! =\! 8.0$, slice volumes $V_2\! =\! 16k,32k$) and in the de Sitter phase ($k_0\! =\! 0.0,5.0$, slice volume $V_2\! =\! 32k$). The error bars correspond to the 95\% confidence intervals of a $\chi^2$-test. }
	\label{fig:haus-micro-width}
\end{figure}

To investigate the influence of the fitting range in a systematic way, 
we have performed fits for a set of ranges $r\!\in\! [r_\textrm{min}, r_\textrm{min}\! +\! w]$ of varying width $w$, and with $r_\textrm{min}\!\in\! [5,14]$. 
The resulting best fit values for the local Hausdorff dimension $d_h$ as a function of $r_\textrm{min}$ are shown in Fig.\ \ref{fig:haus-micro-width},
for two different widths $w\! =\! 10$ and $12$. 

\begin{figure}[t]
	\centering
	\includegraphics[width=0.45\textwidth]{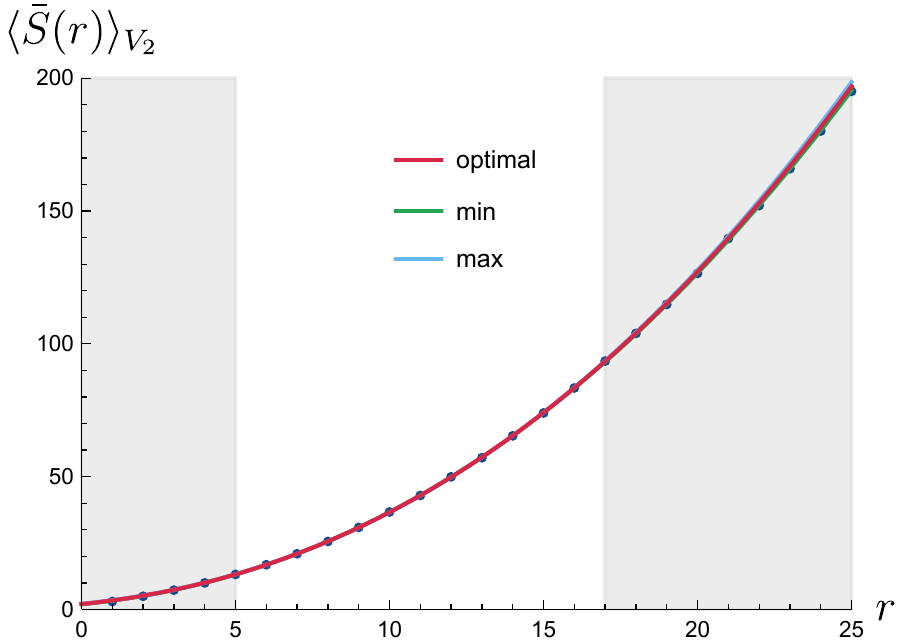}
	\caption{Expectation values $\langle \bar{S}(r)\rangle_{V_2}$ of average shell volumes in the degenerate phase ($k_0\! =\! 8.0$) and slice volume $V_2\! =\! 32k$. The curves are plots of the fit function \eqref{eq:haus-micro-fit} for three different sets of fit parameters: the optimal fit which minimizes $\chi^2$, and the two fits at the boundaries of the 95\% confidence interval. The fit is performed to the data points in the range $5 \leq r \leq 17$, as indicated by the unshaded region.}
	\label{fig:haus-micro-fit-extremes}
\end{figure}

Starting our analysis in the degenerate phase, we observe a good stability of the value of $d_h$ when $r_\textrm{min}$ is increased away from its
lowest, ``canonical'' cutoff value of 5, which indicates that the region $r_\textrm{min}\! \gtrsim\! 5$ is not affected by short-distance artefacts and 
that data in the corresponding interval $r\!\in\! [r_\textrm{min}, r_\textrm{min}\! +\! w]$ are well approximated by a pure power law. Shifting the fitting
range to start beyond $r_\textrm{min}\! =\! 7$ changes the extracted best $d_h$, mildly for the larger volume $V_2\! =\! 32k$ and more strongly for $V_2\! =\! 16k$.
Taking into account Fig.\ \ref{fig:haus-micro-all}, left, this indicates
that one is leaving the region where the functional form (\ref{eq:haus-micro-fit}) is an appropriate fit. Lastly, setting $w\! =\! 12$ instead of 10 reduces
the error bars without appreciably changing $d_h$, and is therefore preferable. To conclude, the optimal choice of range among the possibilities 
we have investigated at $k_0\! =\! 8.0$ appears to be $r\!\in\! [5,17]$. The associated local Hausdorff dimension for $V_2\! =\! 32k$ is given by
\begin{equation}
	d_h = 3.31(4), \quad \quad \textrm{degenerate phase } (k_0=8.0),
\label{dhvalue}
\end{equation}
obtained from fitting to (\ref{eq:haus-micro-fit}), with fit parameters $a\! =\! 4.0(2)$ and $c\! =\! 0.08(1)$.
To illustrate the excellent quality of the fit, Fig.\ \ref{fig:haus-micro-fit-extremes} shows the measured data together with the best-fit curve and the
fits at the edges of the 95\% confidence interval, which basically fall on top of each other 
(see Appendix \ref{app:ci} for details on how to compute such confidence intervals). The fits for the data at $k_0\! =\! 5.0$ and $k_0\! =\! 0.0$
are of similar quality. 
We postpone a discussion of the compatibility of this result with the analytical value $d_h\! =\! 4$ for DT quantum gravity to later, after having
investigated the global Hausdorff dimension.

Turning next to a discussion of the de Sitter phase, Fig.\ \ref{fig:haus-micro-width} shows the best fit values for $d_h$ we extracted from the shell volume data.
They are consistently smaller than those 
in the degenerate phase, and within the de Sitter phase decrease further with decreasing $k_0$. 
We observe only a shorter range of values $r_\textrm{min}\! \gtrsim\! 5$ where the Hausdorff dimension is reasonably stable. 
Examining the corresponding curves in Fig.\ \ref{fig:haus-micro-all}, right, 
there does not seem to be a very extended initial-growth regime, before the curves straighten out to become approximately linear.
It is possible that finite-size effects affect the data points at the upper end of the chosen ranges $[r_{\textrm{min}},r_{\textrm{max}}]$ 
and cause the observed deviations from a simple power-law behaviour. Alternatively, this behaviour may be a bona-fide feature of the embedded slices.
The error bars for larger $r_{\textrm{min}}$ are significantly larger
than in the degenerate phase, especially for $k_0\! =\! 0.0$. At least in part, this appears to be due to  
a statistical uncertainty of the measured shell volumes for increasing $r$.
As before, the error bars are smaller for $w=12$ than for $w=10$. 
For $k_0\! =\! 5.0$, the local Hausdorff dimension seems to decrease slightly with growing $r_{\textrm{min}}$, but in view of the width of the
95\% confidence interval this may not have any significance. 

From a best fit in the interval $r\!\in\! [5,17]$, we find for the local Hausdorff dimension
\begin{align}
	d_h &= 2.91(5), \quad \quad \textrm{de Sitter phase } (k_0=0.0), \\
	d_h &= 3.10(4), \quad \quad \textrm{de Sitter phase } (k_0=5.0),
\end{align}
for the fit parameters $a\! =\! 2.5(3)$, $c\! =\! 0.26(5)$ and $a\! =\! 3.9(3)$, $c\! =\! 0.11(2)$ respectively.
For the time being, we take note of these results and refer to Sec.\ \ref{HDresults} below for a summary and attempted interpretation of
all Hausdorff dimension measurements.

\subsubsection{Global Hausdorff dimension}

The global Hausdorff dimension of a two-dimensional triangulation describes its behaviour as a whole and can again be characterized by the average volume 
$\bar{S}_{N_2}(r)$ of spherical shells of radius $r$, where we have added an explicit subscript $N_2$ to emphasis that the total volume of the triangulation will
now play an important role. Given an ensemble of geometries of volume $N_2$, a global Hausdorff dimension $d_H$ can be extracted if in the limit of large $N_2$
the eigenvalue of the distribution of shell volumes over the entire $r$-range can be described by the functional form
\begin{equation}
\langle \bar{S}_{N_2}(r)\rangle =N_2^{1-1/d_H} \mathcal{F}(x),  \quad \quad x = \frac{r}{N_2^{1/d_H}},
\label{eq:haus-global}
\end{equation}
where $\mathcal F$ is a universal function that depends on the rescaled geodesic distance $x$. 
This is known to be the case for DT quantum gravity in two dimensions, where $\mathcal F$ has been computed explicitly \cite{ambjorn1995fractal},
but there is no guarantee that the scaling law (\ref{eq:haus-global}) holds for general systems of geometries in $D\! =\! 2$. 
Even when a global Hausdorff dimension $d_H$ can
be assigned in this manner, it need not be equal to the local Hausdorff dimension $d_h$ \cite{ambjorn1995fractal,ambjorn1990summing}.

We will attempt to extract a global Hausdorff dimension for the spatial slices in three-dimensional CDT 
by performing a finite-size scaling analysis where we collect data for $\langle \bar{S}_{V_2}(r)\rangle$ for the full range of radii $r$ 
and several slice sizes $V_2$, and then try to rescale the resulting distributions according to \eqref{eq:haus-global}. 
If we can find a single value $d_H$ such that the rescaled distributions fall on top of each other for all volumes $V_2$, we define this $d_H$ 
to be the global Hausdorff dimension of the system.

Following \cite{ambjorn1995fractal}, we will work with the normalized shell volume distributions $n_{V_2}(r)\! :=\! \langle \bar{S}_{V_2}(r)\rangle /V_2$,
for which the scaling law \eqref{eq:haus-global} assumes the form
\begin{equation}
n_{V_2}(r) = V_2^{-1/d_H} \mathcal{F}(x),  \quad \quad x = \frac{r}{V_2^{1/d_H}}.
\label{normal}
\end{equation}
Note that the measured distributions $n_{V_2}(r)$ are functions of a discrete variable $r \in \mathbb{N}_0$. To perform a smooth rescaling, we first construct continuous functions that interpolate between these discrete values, which by slight abuse of notation we continue to call $n_{V_2}(r)$. 
Following the methodology of \cite{barkley2019precision}, we then rescale, for each system volume $V_2$ separately,
the corresponding distribution $n_{V_2}(r)$ such that it maximally overlaps with the normalized distribution $n_{V_\textrm{max}} (r)$ for the largest slice volume $V_\textrm{max}$ in the simulation, which we are using as a reference distribution.
We denote these rescaled distance profiles by $\tilde{n}_{V_2}(\tilde{r})$, where $\tilde{r}$ is a rescaled length variable. They take the form
\begin{equation}
	\tilde{n}_{V_2}(\tilde{r}_{V_2}) =  \left(\frac{V_2}{V_\textrm{max}}\right)^{1/d} n_{V_2}(\tilde{r}), \quad \quad 
	\tilde{r}_{V_2} = \left(\frac{V_2}{V_\textrm{max}}\right)^{1/d}(r+a)-a,
\label{averaging}
\end{equation}
where the two fit parameters are a rescaling dimension $d$ and a phenomenological shift $a$ \cite{ambjorn1995fractal,barkley2019precision} 
that corrects for discretization effects at small $r$, similar to the prescription (\ref{eq:haus-micro-fit}) we used for the local Hausdorff dimension.
\begin{figure}[t]
	\centering
	\includegraphics[width=0.45\textwidth]{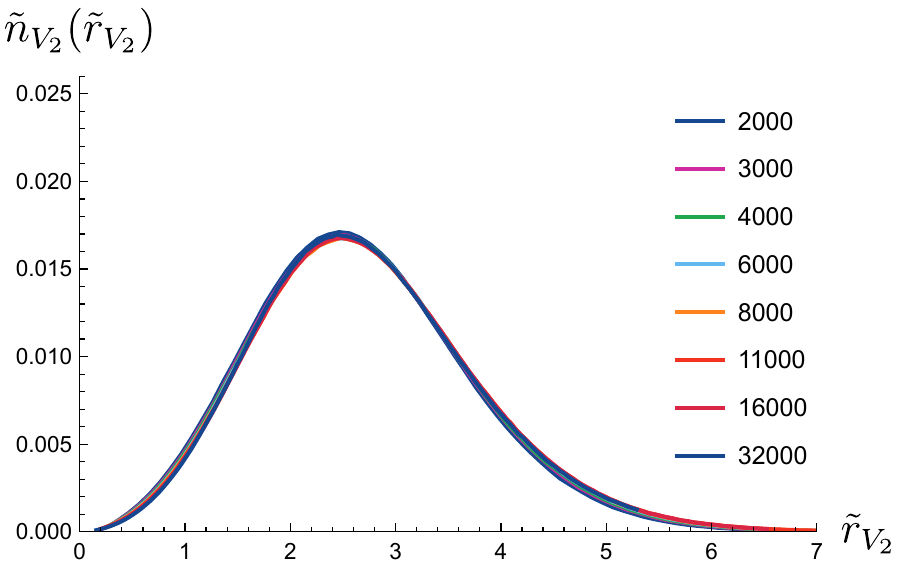} 
	\\
	\includegraphics[width=0.45\textwidth]{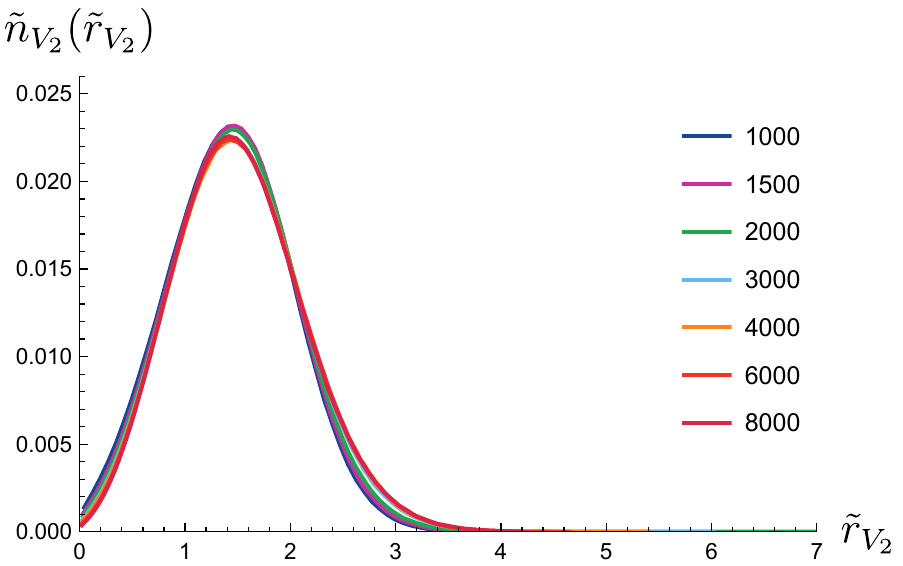}
	\hfill
	\includegraphics[width=0.45\textwidth]{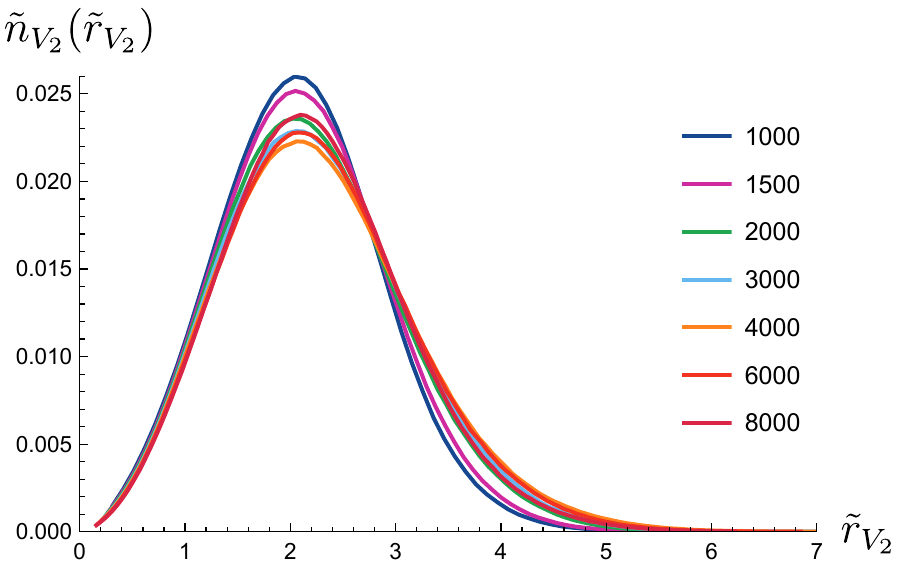}
	\caption{Distributions $\tilde{n}_{V_2}(\tilde{r}_{V_2})$ of shell volumes rescaled with the averaged parameters $(\bar{d},\bar{a})$ 
	in the degenerate phase ($k_0\! =\! 8.0$, top) and the de Sitter phase 
	($k_0\!=\! 5.0$, bottom left, and $k_0\!=\! 0.0$, bottom right).}
	\label{fig:haus-global-collapse}
\end{figure}
Note that $\tilde{r}_{V_\textrm{max}}$ is equal to the original discrete length parameter $r$, so we can use them interchangeably, 
and that $\tilde{n}_{V_\textrm{max}}\! =\! n_{V_\textrm{max}}$.  For each system size $V_2\! <\! V_\textrm{max}$ we determine the corresponding fit parameters 
$a$, $d$ from the condition that the sum of the squared differences $\left(\tilde{n}_{V_2}(\tilde{r}_{V_2})-\tilde{n}_{V_\textrm{max}}(r) \right)^2$ 
should be minimized, where $\tilde{r}_{V_2}$ depends implicitly on the discrete parameter $r$. 
We perform the fit in the range of integers $r$ where $\tilde{n}_{V_\textrm{max}}(r) > \frac{1}{5}\, \textrm{max}_{r}\,\tilde{n}_{V_\textrm{max}}(r)$, 
which means we disregard contributions from the tails of the distribution, where discretization effects are more likely to be present.
If over a large range of $V_2$ the rescaled distributions overlap to a common curve or are reasonably close to doing so, 
we take the mean $\bar{d}$ of all the rescaling dimensions $d$ and define this to be the global Hausdorff dimension, $d_H\! :=\! \bar{d}$.
We also average the shift parameters to obtain one optimal shift $\bar{a}$ for the system. 
If we do not find sufficient overlap, the method fails and we cannot assign a global Hausdorff dimension. 

We measured the expectation value of the shell volume distribution for eight different slice volumes $V_2\! \in\! [1.000,32.000]$ in the degenerate phase, and for seven slice volumes $V_2\!\in\! [1.000,8.000]$ in the de Sitter phase. Again, the autocorrelation times are much larger in the de Sitter phase, and the resulting uncertainties especially large near the peaks of the distributions. Since the location and height of these peaks are important features for finding the appropriate rescaling dimension required for a collapse, large uncertainties in this region imply large error bars for the best fit parameters. 
This led to our choice for the smaller volume range in the de Sitter phase.

\begin{figure}[t]
	\centering
	\includegraphics[width=0.45\textwidth]{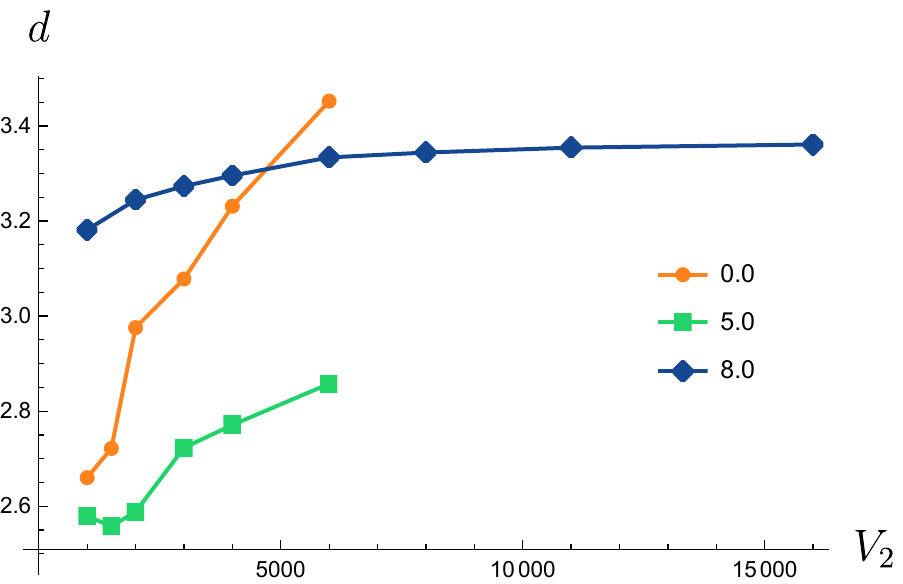}
	\caption{Values for $d$ extracted by comparing measured distributions of shell volumes at a given volume $V_2$ with those at the top volume,
	for $k_0\! =\! 0.0$, $5.0$ and $8.0$, as explained in the text.}
	\label{fig:dcomparison}
\end{figure}

For each of the three phase space points, Fig.\ \ref{fig:haus-global-collapse} shows a collection of curves for a range of volumes $V_2$, which have
been rescaled according to (\ref{averaging}), using the averaged pair $(\bar{d},\bar{a})$ for all of them.  
The topmost graph, for $k_0\! =\! 8.0$, makes a convincing case for the presence of finite-size scaling in the
degenerate phase. Using $\bar{d}\! =\! 3.30$ and $\bar{a}\! =\! 2.64$ as the joint rescaling parameters leads to a curve collapse of good, although not
perfect quality for slice volumes up to $32k$. 
This is not the case for the rescaled curves in the de Sitter phase (Fig.\ \ref{fig:haus-global-collapse}, bottom), which were obtained for
the averaged fit parameters $\bar{d}\! =\! 2.68$ and $\bar{a}\!=\! -0.46$ at $k_0\! =\! 5.0$ and $\bar{d}\! =\! 3.02$ and $\bar{a}\! =\! 2.08$ at $k_0\! =\! 0.0$. 
Although the slice volumes span a significantly smaller range than in the degenerate phase, our rescaled data do not support the presence of finite-size scaling
in this phase. In addition to the mismatch among the curves, we also note that the distributions in the two phases look different as a function of
the rescaled radius $x\! =\! r/V_2^{1/\bar{d}}$. While in the degenerate phase its range extends to around 6.7, and the peak is located near 2.4, 
the slice geometries in the de Sitter phase have a smaller linear extension, with $x$ reaching on the order of 4 (5.7) and the peak located near 1.4 (2.0) 
for $k_0\! =\! 5.0$ (0.0).
One could be tempted to disregard the lack of overlap between the curves for different volumes and simply \emph{define} the global Hausdorff 
dimension to be equal to the average $\bar{d}\! =\! 2.68$ for $k_0\! =\! 5.0$ and $\bar{d}\! =\! 3.02$ for $k_0\! =\! 0.0$. 
However, Fig.\ \ref{fig:dcomparison} shows that this would be misguided, since the values for $d$ 
exhibit a strong dependence on the volume in the range we have investigated. Especially the curve for $k_0\! =\! 0.0$ shows a steep rise, with
little indication of asymptoting to a constant value, very different from the curve for the degenerate phase, which we have included for comparison. 
Note also that the $d$-values extracted from measurements in the de Sitter phase are not in any obvious way related to the values we found for the
local Hausdorff dimension, namely, $d_h\! =\! 3.10(4)$ for $k_0\! =\! 5.0$ and $d_h\! =\! 2.91(5)$ for $k_0\! =\! 0.0$.          
This appears to be yet another indication that the behaviour of the shell distributions is not governed by a single scale and that the scaling hypothesis
(\ref{eq:haus-global}) is simply not valid in the de Sitter phase. 

Based on these observations, we are unable to associate a global Hausdorff dimension with the phase space points in the de Sitter phase.
By contrast, finite-size scaling is observed in the degenerate phase, and the associated global Hausdorff dimension is given by  
\begin{equation}
	d_H = 3.30(2), \quad \quad \textrm{degenerate phase } (k_0 = 8.0),
\end{equation}
which within error margins coincides with the local Hausdorff dimension $d_h\! =\! 3.31(4)$ 
of eq.\ (\ref{dhvalue}) we determined in the previous subsection.  

\subsubsection{Discussion of results}
\label{HDresults}

Let us consider first the results we obtained in the degenerate phase. We found clear evidence for the presence of finite-size scaling when analyzing
the global Hausdorff dimension, and mutually consistent values for both Hausdorff dimensions, with $d_h\! =\! 3.31(4)$ for the local and $d_H\! =\! 3.30(2)$ 
for the global variant. Despite the large discrepancy with the analytical value $d_h\! =\! d_H \! =\! 4$ for two-dimensional DT quantum gravity, we
nevertheless believe that the observed Hausdorff dimension of the spatial slices is compatible with this model of quantum gravity. 
This interpretation is supported by known difficulties in numerically extracting the Hausdorff dimension in two-dimensional systems 
of random geometry (see \cite{loll2015locally} and references therein), with
a tendency of the Hausdorff dimension measurements to underestimate its true value.\footnote{More precisely, this is true for DT models; 
numerical measurements of the Hausdorff dimension for two-dimensional CDT found $d_H\!\approx\! 2$ \cite{ambjorn1999new} and $d_H\! =\! 2.2(2)$ \cite{loll2015locally}.} 
In previous works, these difficulties have motivated the use of a
more general geometric ensemble, the introduction of additional minbu moves and of a phenomenological shift parameter, as well as refined fitting
techniques \cite{ambjorn1995fractal,catterall1995scaling,barkley2019precision}. 
As mentioned earlier, several of these improvements are unfortunately not directly applicable in our case, because our two-dimensional geometries 
are parts of larger, three-dimensional triangulations. 

Earlier numerical results for pure DT quantum gravity whose derivation most closely
resembles our treatment are the finite-size scalings obtained for $N_2\!\leq\! 32k$ from collapsing curves for the shell volume distributions in terms of the 
dual link distance in \cite{catterall1995scaling}. Depending on the fitting 
method, they yielded the values $d_H\! =\! 3.150(31)$ and $d_H\! =\! 3.411(89)$. Unlike ours, this work used a generalized ensemble, but the
final results for the Hausdorff dimension are broadly in line with our findings. Another aspect well illustrated by \cite{catterall1995scaling} is the
increase of the measured Hausdorff dimension with the system volume, something we also observed in the degenerate phase (Fig.\ \ref{fig:haus-micro-width}).

Our results in the de Sitter phase are less clear-cut, but point to a system that is in a different universality class from DT quantum gravity. 
The values we determined for the local Hausdorff dimensions, $d_h\! =\! 3.10(4)$ for $k_0\! =\! 5.0$ and  $d_h\! =\! 2.91(5)$ for $k_0\! =\! 0.0$,
are even further removed from 4, but this might conceivably still be due to some even more serious underestimate than in the degenerate phase.
More significant is the absence of finite-size scaling and the ensuing impossibility to associate a consistent global Hausdorff dimension to the
system. The most likely explanation is the absence of a single scale governing the dynamics, which would imply a different universality class from
that of the degenerate phase. 
Discarding an interpretation of the de Sitter phase in terms of DT quantum gravity leaves us without any obvious alternative candidate theory to explain these results. 
There are a number of two-dimensional CDT models with matter coupling that have a local and/or global Hausdorff dimension of or near 3,
including CDT with eight Ising spins \cite{ambjorn2000crossing}, several massless scalar fields \cite{ambjorn2012pseudotopological},
or restricted hard dimers \cite{ambjorn2014restricted}. Also two pure-gravity models of so-called locally causal dynamical triangulations,
generalizing the strict slicing of two-dimensional CDT, were found to have Hausdorff dimensions near 3 \cite{loll2015locally}. 
There is nothing obvious that would link one of these models to the Euclidean slices
we are dealing with here, but we cannot exclude this possibility either without a further investigation, which however would take us beyond the scope of the
present work.

\subsection{Curvature profile}

The last quantity we will measure on the spatial hypersurfaces is a curvature observable. It is based on the quantum Ricci curvature, a generalized 
notion of Ricci curvature introduced in \cite{klitgaard2018introducing}. The quantum Ricci curvature depends on a neighbourhood of linear size $\delta$ of a
point $x$ and has the interpretation of a coarse-grained Ricci curvature associated with the scale $\delta$. It is defined on a range of metric spaces of
Riemannian signature, including nonsmooth ones, and its introduction was motivated by the search for a notion of (renormalized) curvature suitable for nonperturbative quantum gravity. It can also be defined on classical, smooth Riemannian spaces, where in the limit $\delta\! \rightarrow\! 0$ it
reproduces the standard notion of Ricci curvature \cite{klitgaard2018introducing,klitgaard2018implementing}. The simplest way to turn this
(quasi-)local notion of curvature into a quantum observable depending on the scale $\delta$, dubbed the \emph{curvature profile} \cite{brunekreef2021curvature}, 
is by averaging it over all points of a given metric space and in each point over all directions, leading to a coarse-grained, averaged notion of 
a Ricci scalar. The quantum Ricci curvature and associated curvature profiles have been studied for a wide range of smooth and piecewise flat spaces
\cite{klitgaard2018implementing,brunekreef2021curvature} and used to characterize the curvature properties of 
DT quantum gravity in $D\! =\! 2$ \cite{klitgaard2018implementing}, CDT quantum gravity in $D\! =\! 2$ \cite{brunekreef2021quantum}, and full,
four-dimensional CDT quantum gravity, producing further evidence for the de Sitter nature of the quantum geometry in four dimensions \cite{klitgaard2020how}. 
In the following, we will recall briefly the ingredients and construction of the curvature profile, and refer to the literature \cite{klitgaard2018introducing,brunekreef2021curvature,brunekreef2021quantum} for more detailed discussions.

\begin{figure}[t]
	\centering
	\includegraphics[width=0.45\textwidth]{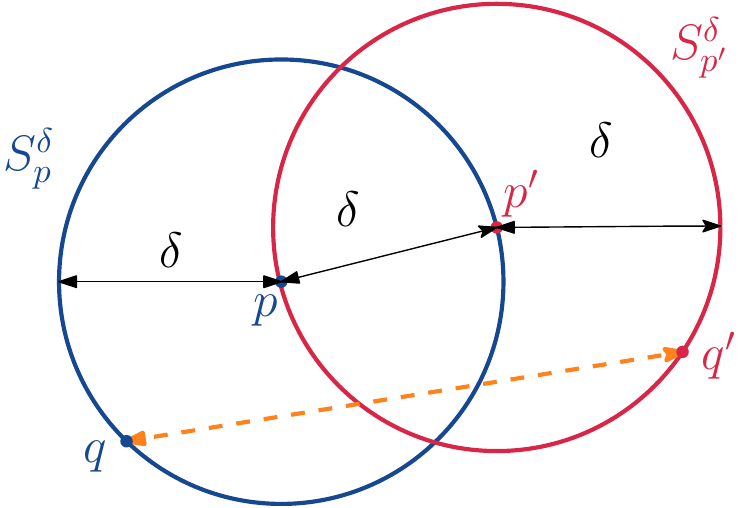}
	\caption{The average sphere distance $\dbar$ of two geodesic circles $S_p^\delta$ and $S_{p'}^\delta$, whose centres $p$ and $p'$ are a distance 
	$\delta$ apart, is obtained by averaging over the distance $d(q,q')$ of all point pairs $(q,q')$ along the two circles.}
	\label{fig:avg-sphere-dist}
\end{figure}

The main ingredient in determining the quantum Ricci curvature is the measurement of the \emph{average sphere distance} $\bar{d}(S_p^\delta, S_{p'}^\delta)$ 
of two overlapping geodesic spheres $S_p^\delta$, $S_{p'}^\delta$, each of radius $\delta$, whose centres $p$ and $p'$ are also a geodesic distance $\delta$ apart. To compute the ($\delta$-dependent) quantity $\bar{d}$, we average over the distance between all pairs of points 
$(q,q') \in S_p^\delta \times S_{p'}^\delta$ on 
the two spheres. The geometry of the situation is depicted in Fig.\ \ref{fig:avg-sphere-dist} for the two-dimensional case, where the spheres are given by 
circles. When applying the prescription on a smooth Riemannian manifold, one extracts the Ricci curvature $Ric(v,v)$ at $p$ in the limit $\delta\!\rightarrow\! 0$,
where $v$ is the tangent vector at $p$ in the direction of $p'$ \cite{klitgaard2018introducing,klitgaard2018implementing}. 
However, we will implement the prescription on piecewise flat triangulations and for non-infinitesimal $\delta$, where distances are
defined in terms of an integer-valued link distance or dual link distance, and the limit $\delta\!\rightarrow\! 0$ is neither well defined nor physically interesting because of short-distance discretization artefacts.   
In what follows, we will use the link distance $d(q,q')$ between pairs $q,q'$ of vertices, to allow for a direct comparison with previous measurements 
of the quantum Ricci curvature in two-dimensional DT quantum gravity, which also used the link distance \cite{klitgaard2018implementing}, on the
ensemble of regular, simplicial manifolds.
In the triangulated setting, a geodesic ``sphere'' $S_p^\delta$ centred at the vertex $p$ is defined as the set of vertices at link distance $\delta$ from $p$, 
$N_0(S_p^\delta)$ counts the number of vertices in this set, and the average sphere distance takes the form of a normalized double sum
\begin{equation}
	\dbar (S^\delta_{p},S^\delta_{p'} ) = \frac{1}{N_0(S^\delta_{p})} \frac{1}{N_0(S^\delta_{p'})} \sum_{q \in S^\delta_{p}} \sum_{q' \in S^\delta_{p'}} d(q,q'), 
	\;\;\;\;\; d(p,p')=\delta .
	\label{eq:disc-avg-sph-dst}
\end{equation}
We have used inverted commas since the vertices of a point set $S_p^\delta$ will in general not form a genuine sphere: 
it is not required that the vertices form a sequence of nearest neighbours 
which can be joined pairwise by $N_0(S_p^\delta)$ edges, resulting in a unique one-dimensional simplicial submanifold of the topology of a circle. 
Rather, such a procedure generally yields multiple circles, and results in self-intersections and -overlaps.

The quantum Ricci curvature $K_q(p,p')$ associated with the point pair $(p,p')$ is defined in terms of the normalized average sphere distance as
\begin{equation}
\bar{d}(S_p^{\delta},S_{p'}^{\delta})/\delta=:c_q\, (1 - K_q(p,p')), \quad \quad \delta = d(p,p').
\label{eq:qric}
\end{equation}
The factor $c_q$ is a positive constant which describes the $\delta$-independent part of the average sphere distance. In the continuum, it can be defined 
by the limit $c_q \!:=\! \lim_{\delta \to 0} \dbar / \delta$ and depends only on the dimension of the manifold. The function $K_q(p,p')$ captures the non-trivial dependence of the average sphere distance on the direction of the vector $\overline{p p'}$ and the scale $\delta$. 
The most straightforward way to construct a genuine, diffeomorphism-invariant curvature observable is by taking an average $\bar{d}_{\rm av}$
of the average sphere distance $\bar{d}$ of eq.\ (\ref{eq:qric}) over all pairs $(p,p')$ of centre points at a fixed distance $\delta = d(p,p')$ in the
triangulation $T$,
\begin{equation}
\bar{d}_{\rm av}  (\delta):	=  \frac{1}{{\cal N}_\delta} \sum_{p \in T}\sum_{p' \in T} \dbar (S_p^\delta, S_{p'}^\delta) \, \delta_K(d(p,p'), \delta).
	\label{eq:dbar-av}
\end{equation}
The symbol $\delta_K$ denotes a discrete Kronecker delta, implementing the distance constraint on $p,p'$, and the normalization ${\cal N}_\delta$ 
is defined by the double sum
\begin{equation}
	{\cal N}_\delta = \sum_{p \in T}\sum_{p' \in T}  \delta_K(d(p,p'), \delta).
\end{equation}
The so-called curvature profile, introduced in \cite{brunekreef2021curvature}, is given by the quotient $\bar{d}_{\rm av}(\delta)/\delta$.
Since the double sum \eqref{eq:dbar-av} includes an average over all directions, it allows us to extract a scale-dependent 
quantum Ricci scalar $K_{\rm av}(\delta)$ from the curvature profile via
\begin{equation}
\bar{d}_{\rm av}(\delta)/\delta=:c_{\rm av} (1 - K_{\rm av}(\delta)).
\label{eq:curv-prof}
\end{equation}
Since in the simplicial setting the factor $c_\textrm{av}$ cannot be fixed through a limit $\delta\! \to\! 0$, it is set to the expectation value of the 
normalized average sphere distance for the minimal value of $\delta$ such that discretization artefacts are no longer dominant. 
In \cite{klitgaard2018implementing}, this value was taken to be $\delta\! =\! 5$. 
\begin{figure}[t]
	\centering
	\includegraphics[width=0.65\textwidth]{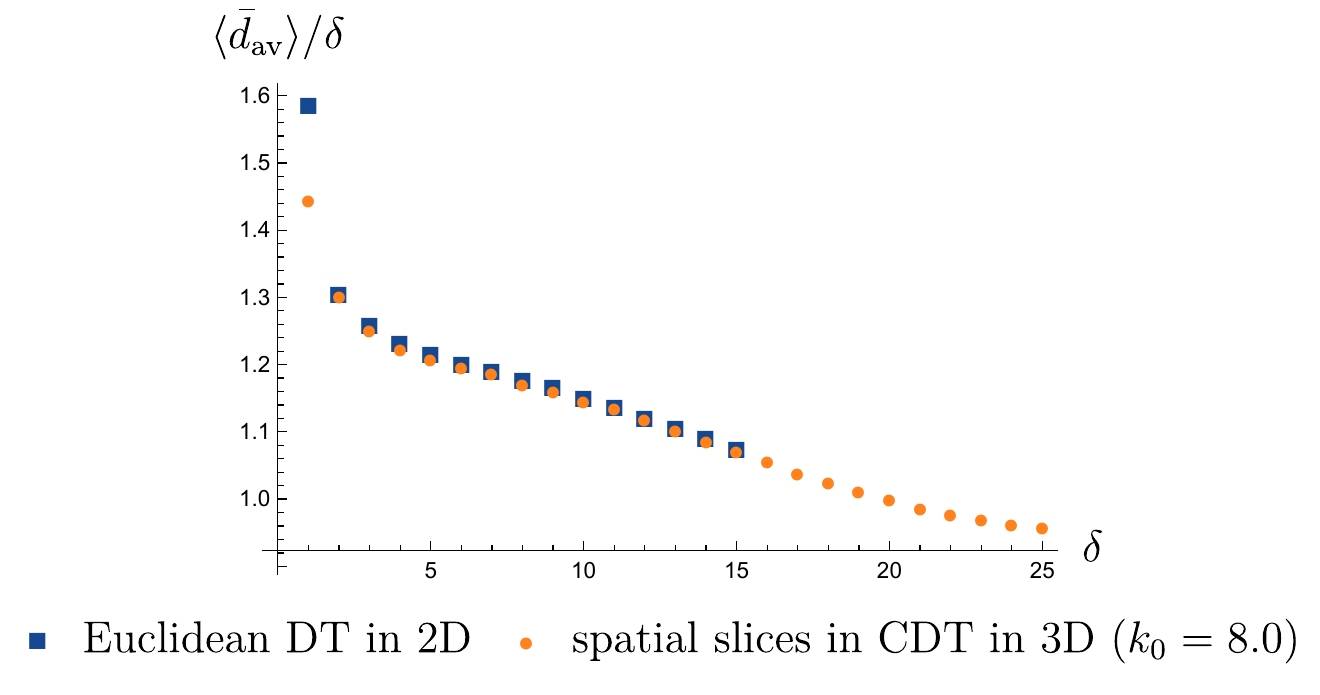}
	\caption{Expectation value $\langle \dbar_{\rm av} / \delta \rangle$ of the normalized average sphere distance, measured in two-dimensional 
	DT quantum gravity (blue squares) \cite{klitgaard2018implementing}, 
	and for the spatial slices of three-dimensional CDT quantum gravity in the degenerate phase (yellow dots), both for volume $V_2\! =\! 60k$. 
	(Error bars are smaller than dot size.)}
	\label{fig:ricci-60k}
\end{figure}

Continuing our investigation of three-dimensional CDT quantum gravity, we measured the expectation value 
$\langle \dbar_{\rm av} / \delta \rangle$ of the curvature profile of its spatial hypersurfaces at the phase space points $k_0\! =\! 0.0$, 5.0 and 8.0. 
We used slice volumes in the range $V_2\!\in\! [4k, 60k]$ in the degenerate phase and $V_2\!\in\! [4k,20k]$ in the de Sitter phase.
For the volumes $V_2\! =\! 20, 30, 40$ and $60k$, we compared the curvature profiles in the degenerate phase at $k_0\! =\! 8.0$ to those of
two-dimensional DT quantum gravity \cite{klitgaard2018implementing}, and in each case found agreement within statistical error bars \footnote{We thank N.\ Klitgaard for making the original data available to us.}. 
For illustration, the results for $V_2\! =\! 60k$ are shown in Fig.\ \ref{fig:ricci-60k}.
There is an excellent match of our present data, taken for $\delta\! \in\! [1, 25]$, with those of DT quantum gravity in the range where they overlap,
except at $\delta\! =\! 1$. The fact that this agreement extends even into most of the region of discretization artefacts at small $\delta$ 
provides additional evidence that the spatial hypersurfaces in the degenerate phase behave like two-dimensional DT geometries. 

The monotonically decreasing curvature profiles we found in both the degenerate and the de Sitter phase clearly indicate the presence of 
positive curvature, as can be seen from eq.\ \eqref{eq:curv-prof}. 
Motivated by the fact that curvature profiles of two-dimensional DT quantum gravity can best be fitted to those of a five-dimensional continuum sphere
with some effective curvature radius $\rho_\textrm{eff}$ \cite{klitgaard2018implementing}, we tried to do the same for our data. 
As expected from the match of the curvature profiles, our results for the effective curvature radii in the degenerate phase are close 
to the values listed in Table 1 of \cite{klitgaard2018implementing}. Note that a rough estimate for the onset of 
finite-size effects in measuring $\langle \dbar_{\rm av} / \delta \rangle$ on a sphere of curvature radius $\rho$ is $\delta\!\approx\rho$, where 
the extension $3\delta$ of the double circle of Fig.\ \ref{fig:avg-sphere-dist} is 
approximately equal to $\pi\rho$, half of the circumference of the sphere. This is in good agreement with the findings 
in \cite{klitgaard2018implementing}.

\begin{figure}[t]
	\centering
	\includegraphics[width=0.45\textwidth]{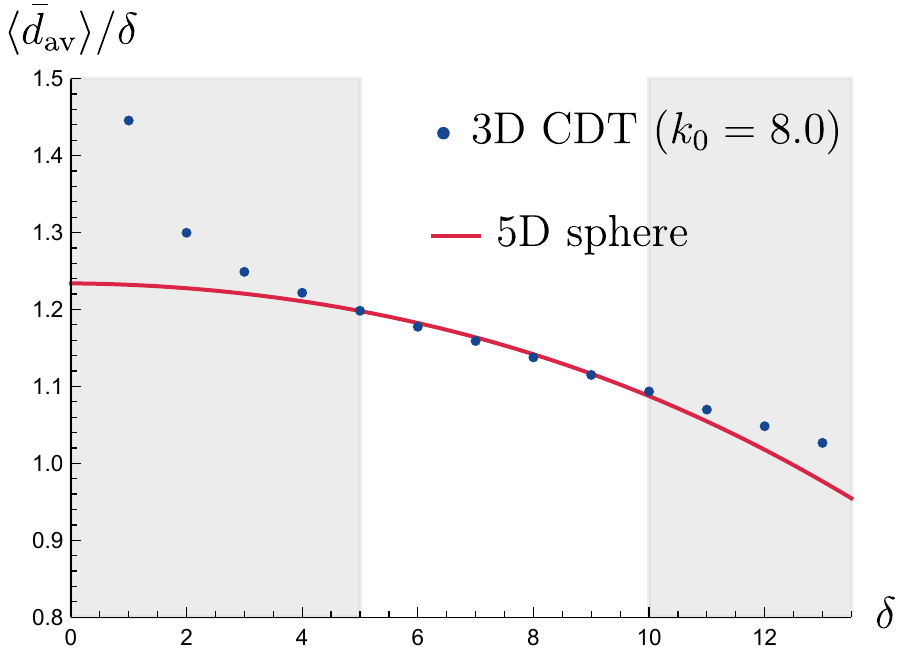}
    \hfill
    \includegraphics[width=0.45\textwidth]{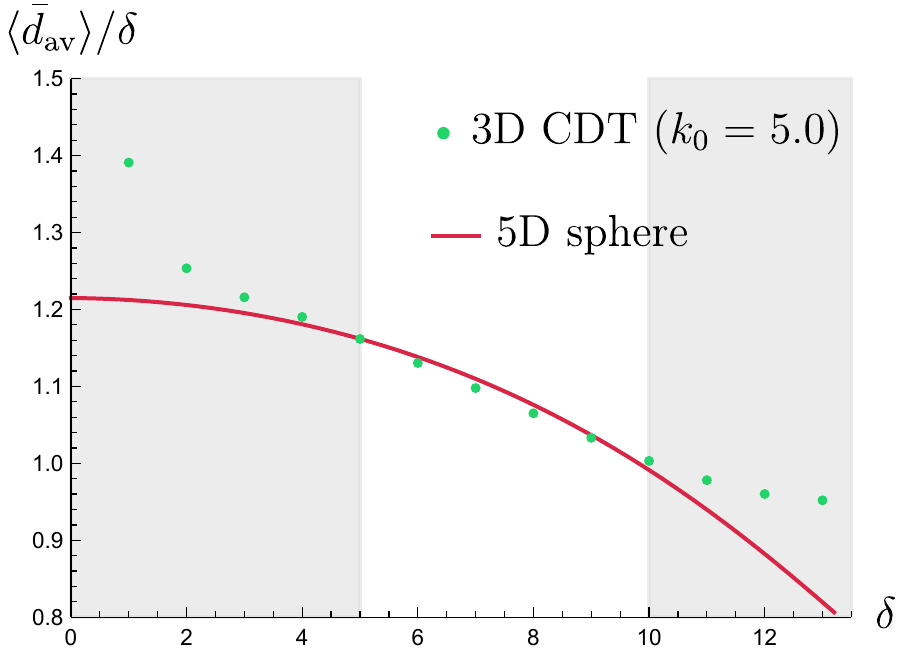}
	\caption{Comparing measured curvature profiles in the range $\delta\! \in\! [5, 10]$ with those 
	of five-dimensional continuum spheres, for slice volume $V_2\! =\! 20k$. Left: fit to a sphere with $\rho_\textrm{eff}\! =\! 13.5$ in the degenerate phase 
	($k_0\! =\! 8.0$). Right: fit to a sphere with $\rho_\textrm{eff}\! =\!11.1$ in the de Sitter phase ($k_0\! =\! 5.0$).}
	\label{fig:curvfits}
\end{figure}

Consistent with this argument, we found that for $V_2\! =\! 60k$, a fitting range $\delta\!\in\! [5,15]$ is appropriate,
while for the smaller slice volume $V_2\! =\! 20k$, which is the maximal size available for measurements in the de Sitter phase, 
the smaller range $\delta\!\in\! [5,10]$ should be used. 
The measured curvature profile for $V_2\! =\! 20k$ at $k_0\! =\! 5.0$ in the de Sitter phase is shown in Fig.\ \ref{fig:curvfits}, 
where for comparison we present it alongside 
the result for the same slice volume at $k_0\! =\! 8.0$ in the degenerate phase. 
The continuous lines are best fits to a 5D continuum sphere. Following \cite{klitgaard2018implementing}, an additive shift was used such that the 
data point at $\delta\!=\!5$ always lies on the continuum curve. 
Because of the small fitting range we cannot and do not claim that the data taken at this (or even smaller) volume represent convincing evidence for
the curvature behaviour of a sphere, and the effective curvature radii extracted ($\rho_\textrm{eff}\! =\!13.5$ for $k_0\! =\! 8.0$, 
$\rho_\textrm{eff}\! =\!11.1$ for $k_0\! =\! 5.0$) should be taken with a large grain of salt.\footnote{The data at $k_0\! =\! 0.0$ are somewhat similar
to those at $k_0\! =\! 5.0$, but their quality is even worse, and we do not show them here.}   
In the degenerate phase we can say a bit more, as we have seen, since we can go up to volume $V_2\! =\! 60k$ and essentially reproduce the
results of DT quantum gravity. With regard to the de Sitter phase, we can at this stage only conclude that the curvature profiles are not in contradiction 
with those of a 5D continuum sphere, but it is clear that the quality of our results is not sufficient to definitely say that it is a sphere, let alone determine
its dimension and curvature radius reliably. This would require us to probe much larger systems, which especially in the de Sitter phase was not feasible
in our set-up.

\section{Summary and conclusion}
\label{sec:disc}

We set out to gain a more detailed understanding of the properties of three-dimensional CDT quantum gravity by studying the intrinsic geometric properties of its
spatial slices at integer proper time. We worked with the ``classic'' ensemble of three-dimensional simplicial manifolds, 
which implies that the spatial hypermanifolds under
consideration also satisfy manifold conditions, and can be characterized by dual trivalent graphs without tadpoles or self-energy insertions.  
The original work on this quantum gravity model found two distinct phases on either side of a first-order transition as a function of the bare inverse
gravitational coupling $k_0$ \cite{ambjorn2001nonperturbative}: a de Sitter phase of extended geometry for $k_0\! <\! k_0^c$ and a degenerate phase
for $k_0\! >\! k_0^c$, characterized by a strongly fluctuating volume profile, the almost complete absence of (2,2)-tetrahedra and an approximate
decoupling of nearby spatial slices. 

We investigated the intrinsic geometric properties of the spatial slices in both phases, well away from the critical coupling $k_0^c\!\approx\! 6.24$,
at $k_0\! =\! 0.0$ and $5.0$ in the de Sitter and at $k_0\! =\! 8.0$ in the degenerate phase. The quantities considered were the expectation values
of the average coordination number of vertices in the slices, of the entropy exponent extracted from the distribution of minimal-neck baby universes, of the local 
and global Hausdorff dimension, and of the curvature profile, obtained by averaging the quantum Ricci curvature. They are for the most part well studied in
two-dimensional DT and CDT quantum gravity, the primary systems of reference for our results on the spatial slices. 
One aim of our investigation, motivated by 
the observed decoupling behaviour of the spatial slices, was to verify that the behaviour of the slices in the degenerate phase lies in the same
universality class as DT quantum gravity in $D\! =\! 2$. What happens in the de Sitter phase, and to what extent the embedding three-dimensional 
geometry influences the effective dynamics of the hypersurfaces in this phase is much less clear a priori.

Summarizing our results, we found convincing evidence that the behaviour of the spatial slices in the degenerate phase is indeed compatible with that of
two-dimensional DT quantum gravity. The measured distribution of the vertex order follows the analytical prediction almost perfectly 
(Fig.\ \ref{fig:cdt-coord-results-nonlog}). The same is true for the distribution of (sufficiently large) minbu sizes (Fig.\ \ref{fig:minbu-dist}, top), yielding
an entropy exponent $\gamma\! =\! -1/2$, the known value for pure Euclidean gravity in $D\! =\! 2$. Measurement of the local and global Hausdorff dimension yielded mutually compatible results, with $d_h\! =\! 3.31(4)$ and $d_H\! =\! 3.30(2)$, exhibiting finite-size scaling for the latter. 
We argued that the discrepancy between the measured values and the analytically known value $d_h\!\equiv\! d_H\! =\! 4$ is in line with expectations for
a simplicial manifold ensemble and the relatively small volumes under consideration here. Finally, the measured curvature profile matched very well that of a
previous investigation of DT quantum gravity in $D\! =\! 2$ (Fig.\ \ref{fig:ricci-60k}), at least up to the slice volume $V_2\! =\! 60k$ we could investigate.

By contrast, in the de Sitter phase we could not establish a corresponding overall match of the behaviour of the measured observables with that of any 
known quantum gravity 
model in two dimensions. In particular, we did not find any evidence that the spatial slices behave according to DT quantum gravity in $D\! =\! 2$,
which one may argue is the most natural hypothesis, given the absence in the slices of a preferred direction or a time-space asymmetry, 
which characterizes two-dimensional
CDT configurations. We also found that the behaviour within the de Sitter phase depends on the value of the bare coupling constant $k_0$. 
Since it was tangential to our main focus, we did not study the nature of this $k_0$-dependence more closely, which may be a worthwhile project in itself. 
Within the limited range of couplings $k_0\!\in\! [3.0,6.0]$, earlier work found some evidence that 
results inside the de Sitter phase can be mapped onto each other by a $k_0$-dependent rescaling of the time- and space-like length units \cite{ambjorn2001nonperturbative}. It would be interesting to understand whether this also extends to the value $k_0\! =\! 0$ we have been using,
or even to negative $k_0$.

Returning to the specifics of our results, the vertex order in the de Sitter phase was found to obey a very different distribution from that in the degenerate phase, with large coordination numbers being more prevalent (Fig.\ \ref{fig:cdt-coord-results}). We also saw that the distribution for $k_0\! =\! 5.0$ is even
further removed from that for $k_0\! =\! 8.0$ than the distribution for $k_0\! =\! 0.0$.
The method of determining the entropy exponent $\gamma$ from the distribution of minbu sizes does not appear to be applicable in the de Sitter phase,
which we conjectured to be due to the presence of correlations in the three-dimensional embedding triangulations. Although this does not 
necessarily disprove a DT-like behaviour of the spatial slices, it does not present any evidence in favour of it either. The measured local Hausdorff
dimensions, given by $d_h\! =\! 3.10(4)$ for $k_0\! =\! 5.0$ and $d_h\! =\! 2.91(5)$ for $k_0\! =\! 0.0$ are significantly smaller than the value found in the
degenerate phase, and point to a different continuum limit than that of two-dimensional DT quantum gravity. 
Of course, it could be the case that the de Sitter phase is subject to much larger discretization artefacts, because of the genuinely three-dimensional 
nature of the underlying geometries, and that one needs to go to larger slice volumes to get a better approximation of continuum behaviour. 
However, even taking this possibility into account, a yet stronger indication that
the slices do not exhibit DT-like behaviour comes from the absence of finite-size scaling at fixed $k_0$ to extract a global Hausdorff dimension, from
which we deduced that the slice dynamics is likely governed by more than just one scale. 
Finally, the measurement of the curvature profiles showed the presence of a positive average quantum Ricci scalar. Matching with
a continuum sphere was in principle possible, but should at this stage be regarded as inconclusive, since it was based on only a handful of measurement 
points, which could be compatible with other curvature profiles also. It therefore cannot serve as evidence that the system is equivalent to
two-dimensional DT quantum gravity. 

Having largely dismissed an interpretation of the spatial slices in the de Sitter phase in terms of two-dimensional DT or CDT quantum gravity does not leave 
any obvious alternatives to associate them with (the universality classes of) other known systems of random geometry. 
DT gravity coupled to matter with a conformal charge $c\! <\! 1$ is disfavoured,
because the apparent absence of finite-size scaling for the global Hausdorff dimension of the spatial slices contradicts the scale-invariance of these systems.   
On the one hand, the quality of our data is far removed from the precision measurements of the Hausdorff dimension of such 
systems \cite{barkley2019precision}, 
and we cannot entirely exclude that finite-size scaling will appear at much larger volumes than the ones we could probe here.
On the other hand, we would urge caution when comparing to ensembles with less stringent regularity conditions, like those used in \cite{barkley2019precision}: even if the relaxation of simplicial
manifold conditions does not change the universality class in specific two-dimensional models, this may not hold in general 
in three-dimensional quantum gravity models, and may depend on the details of the regularity conditions.
This is part of a more general question, namely, are there natural larger ensembles of three-dimensional triangulations which contain the 
simplicial manifold ensemble of CDT, but lie in the same universality class? Conversely, are there strictly smaller ensembles contained in that
of standard CDT quantum gravity, which still belong to the same universality class? Larger ensembles may facilitate
numerical simulations and lead to faster convergence, while smaller ensembles may be easier to enumerate and handle analytically (see \cite{budd2022family} for a recent example in three-dimensional DT quantum gravity). 
Of course, there may be more than one physically interesting universality class associated with three-dimensional Lorentzian random geometries, 
like the wormhole phase described by the ABAB-matrix model \cite{ambjorn2001lorentzian} already mentioned in the introduction.

Returning to ensembles of simplicial manifolds, our results in the de Sitter phase may indicate the existence of another, new model of
two-dimensional quantum geometry, where the embedding three-dimensional geometry induces some effective dynamics on the spatial slices,
presumably through $k_0$-dependent extrinsic curvature contributions. Further research is needed to understand whether such an induced
model exists and whether it can in turn be interpreted as a two-dimensional quantum field theory with properties like locality and unitarity, as is
the case in the degenerate phase.
Contrasting our relatively straightforward verification of the DT nature of the spatial dynamics in the degenerate phase with
the difficulties we encountered when investigating the de Sitter phase highlights 
the fact that quantum geometry in three dimensions -- here by leaving its imprint on the hypersurfaces -- is significantly more complex and 
complicated than quantum geometry in two dimensions. Much remains to be done to illuminate its mathematical and physical properties.

\vspace{0.5cm}
\noindent{\bf Acknowledgments.}
This work was partly supported by a Projectruimte grant of the Foundation for Fundamental Research on Matter (FOM, now defunct), financially supported by the Netherlands Organisation for Scientific Research (NWO). J.B.\ would like to thank Wouter van Amsterdam for useful comments on Appendix \ref{app:ci}.

\begin{appendices}
\section{Estimating the entropy exponent}
\label{app:entropy}
This appendix describes the procedure used in Sec.\ 2.1 of \cite{ambjorn1993baby} to determine best fit values for the entropy exponent $\gamma$ 
with associated error bars, together with the results of this analysis applied to our data. One introduces a subleading correction to the right-hand 
side of \eqref{eq:minbu-dist} by replacing
\begin{equation}
	B^{\gamma-2} \to B^{\gamma-2}\left(1+\frac{c}{B}+O\left(\frac{1}{B^2}\right)\right),
\end{equation}
which allows for a better fit in the regime of small $B$, without significantly affecting the behaviour of the function at intermediate and large $B$. 
One then takes the logarithm on both sides of \eqref{eq:minbu-dist} to obtain
\begin{equation}
\label{eq:log-minbu-cor}
	\log(\bar{n}_{N_2}(B)) = a + (\gamma-2)\log\left(B(N_2-B)\right) + \frac{c}{B},
\end{equation}
where the best fit parameters $a$, $\gamma$ and $c$ should now be determined from the observed baby universe distributions shown in 
Fig.\ \ref{fig:minbu-dist}.\footnote{Note that the argument in the logarithm on the right-hand side of this expression differs from eq.\ (2.2) in \cite{ambjorn1993baby} by a multiplicative factor $N_2$, which is absorbed by the fit parameter $a$.} 
The goodness of fit is defined through the $\chi^2$-statistic, described in greater detail in App.\ \ref{app:ci} below.

\begin{figure}[t]
	\centering
	\includegraphics[width=0.45\textwidth]{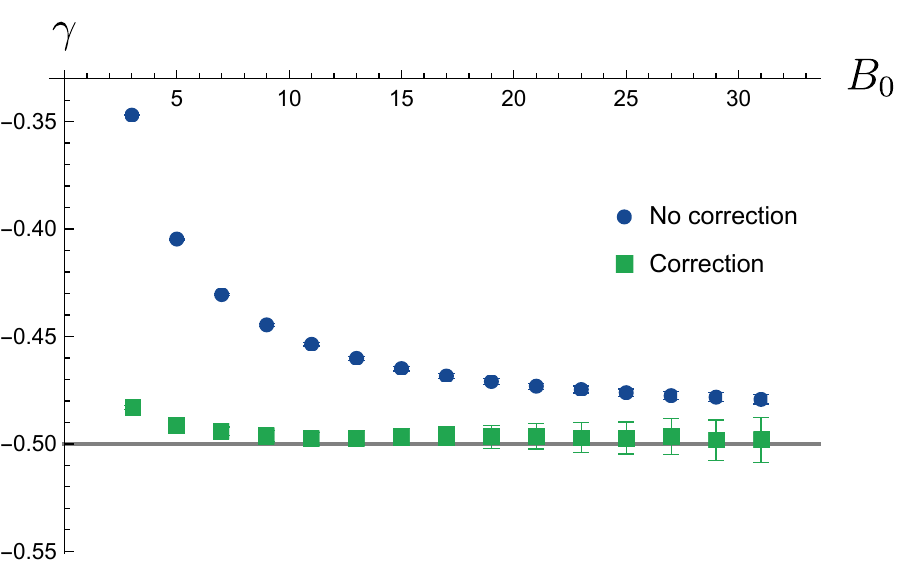}
	\caption{Fitted values of the entropy exponent $\gamma$ in the degenerate phase for slice volume $N_2\! =\! 1.000$ and different lower cut-offs $B_0$. 
	We show the best fit values with and without the correction term appearing in \eqref{eq:log-minbu-cor}.}
	\label{fig:minbu-gamma-fits}
\end{figure}

As mentioned before, the prediction \eqref{eq:minbu-dist} is only expected to hold for sufficiently large baby universes, where discretization effects are negligible. 
We therefore introduce a lower cut-off $B_0$ on $B$ on the data before extracting the best fit parameters. The resulting values of the entropy
exponent $\gamma$ as a function of the cut-off $B_0$ are shown in Fig.\ \ref{fig:minbu-gamma-fits} for $N_2\!\equiv V_2\! =\! 1.000$, both with and without the correction term $c/B$ in \eqref{eq:log-minbu-cor}. Our figure resembles Fig.\ 2 of reference \cite{ambjorn1993baby} extremely closely, including the magnitude of the error bars. 
The authors of \cite{ambjorn1993baby} subsequently obtain an estimate for $\gamma$ by fitting an exponential of the form 
\begin{equation}
	\gamma(B_0) = \gamma - c_1\, e^{-c_2\, B_0}.
\end{equation}
The resulting values for $\gamma$ shown in Table 1 of \cite{ambjorn1993baby} are (within statistical error) identical to the ones we found from our data, using the same method.

\section{Determining confidence intervals for best fit parameters}
\label{app:ci}
While researching the literature on numerical estimates of quantities like the Hausdorff dimension and the entropy exponent, we found that previous work 
is often not very explicit about the methodology used to determine the uncertainty in the best fit parameters, if such margins of error are provided at all. 
Since our goal was to investigate whether the behaviour of the spatial slices is consistent with that of known models of two-dimensional random geometry, 
we considered it important to attach a degree of confidence to the results obtained. This led us to a more detailed study of the statistical methods available for performing such an analysis. Here we summarize our findings, in the hope that others may find them useful. We also comment on the assumptions required to make use of these methods, and to what extent they apply in the present context. Our main aim is to motivate the choices made in computing the confidence intervals; we do not aim for full mathematical or statistical rigour.

The generic starting point consists of a set of $N$ data points $y_n \in \mathbb{R}$ measured at positions $\vec{x}_n$, to which we want to fit a function $f(\vec{x},\vthet)$ 
with parameters $\vthet$. Firstly, we require a measure of the goodness-of-fit which allows us to find the set of best fit parameters $\vthetm$ that minimizes this measure. Secondly, since the measurements are inherently noisy, we are interested in specifying a range for the fit parameters in which we expect to find the ``true'' values with a certain degree of confidence. In the method of least squares, the goodness-of-fit is defined through the sum of squares of the residuals, and the optimal fit is obtained when this sum is minimized. However, the standard unweighted sum of least squares assumes equal variances on all the data points, 
which typically is not the case for the measurements we perform in lattice quantum gravity. In what follows, we will show how the $\chi^2$-distribution can be used in the context of a linear regression model to define a goodness-of-fit and corresponding confidence intervals for the parameters $\vthet$. The models we fit in the main text of this work do not fall into this class of linear models, so we subsequently discuss how the analysis is affected when the condition of linearity is relaxed.

\subsection{Best fit parameter estimation}
A linear model takes the form
\begin{equation}
	f\left(\vec{x}; \vec{\theta}\right) = \sum_{i=1}^p \theta_i f_i(\vec{x}),
\end{equation}
where the $f_i$ are functions of the independent variables. These functions are allowed to be nonlinear in the $x_i$ --- the linearity condition applies to the fit parameters $\theta_i$ only. Suppose the ``correct'' model has fit parameters $\vthet_0$. If the errors $\sigma_n$ on the $N$ measurement outcomes are Gaussian, we can consider the $y_n$ to be normally distributed random variables with mean $f(\vec{x}_n; \vthet_0)$ and standard deviation $\sigma_n$. We can turn these into $N$ standard normals $Z_n=\frac{y_n-f(\vec{x}_n;\vthet_0)}{\sigma_n}$. Let us define the \emph{$\chi^2$-statistic} of a choice of fit parameters $\vthet$ as the sum of the squares of the $Z_n$,
\begin{equation}
\label{eq:chisq}
	\chi^2\left(\vec{\theta}\right) = \sum_{n=1}^N \left(\frac{y_n-f\left(\vec{x}_n; \vthet\right)}{\sigma_n\,}\right)^2.
\end{equation}
The reason we call this the $\chi^2$-statistic is that the sum of $k$ independent standard normals follows a so-called $\chi^2$-distribution with $k$ degrees of freedom. Such a distribution has expectation value $k$ and variance $2k$. The quantity $\chi^2\left(\vthet\right)$ should be minimized to find a \emph{maximum likelihood estimator} for $\vthet_0$, corresponding to the set of best fit parameters.

In the current situation, where we are trying to fit a model to the data points, we must take into account that the individual $y_n$ are explicitly \emph{not} independent --- after all, we have hypothesized the existence of a model $f(\vec{x}; \vthet)$ that can predict the outcomes of our measurements. The sum of $N$ squared standard normal distributions therefore follows a $\chi^2$-distribution with a certain number $k\! <\!  N$ of degrees of freedom. For a linear model without priors (i.e.\ restrictions on the fit parameters), we have $k\! =\! N-p$, where $p$ is the number of fit parameters.

Therefore, when fitting a linear model $f(\vec{x}; \vthet)$ with $p$ fit parameters to $N$ data points $y_n$ with Gaussian errors $\sigma_n$, we expect the $\chi^2$-statistic to follow a $\chi^2$-distribution with $k=N-p$ degrees of freedom. As mentioned earlier, the best fit parameters $\vthetm$ are determined by finding the minimum possible value $\chi^2_\textrm{min}$ of the $\chi^2$-statistic. The fit is considered good when $\chi^2_\textrm{min}\! \approx\! k$, since this is the expectation value of the corresponding $\chi^2$-distribution. Significantly larger values of $\chi^2_\textrm{min}$ indicate that no reasonable fit could be found (e.g.\ our assumptions about the model may be wrong), whereas a $\chi^2_\textrm{min}$ significantly lower than $k$ could mean we are overfitting the data or overestimating the measurement errors $\sigma_n$.

\subsection{Confidence intervals}
With the best fit parameters $\vthetm$ at our disposal, we can now turn to determining \emph{confidence intervals} on these parameters. After all, slightly varying the parameters around their best fit values should produce approximately equal $\chi^2$-statistics. Moreover, the measurement data we are using to compute 
$\chi^2$ is inherently noisy, which implies that we have merely obtained an estimate of the ``true'' best fits. To specify a range in which we expect to find the true values with a certain degree of confidence, we can use the properties of the $\chi^2$-distribution. 

Confidence intervals (CIs) are computed at a certain \emph{confidence level}, specified by a percentage (a common choice is the 95\% CI). Alternatively, we can specify a significance level $\alpha$, corresponding to a $(1\!-\! \alpha)\%$ confidence level. Given $\alpha$, we can determine the \emph{critical value} $\chi^2_{\textrm{crit},\alpha}$ of the $\chi^2$-statistic for our fitted model containing $p$ parameters by solving $P(\chi^2\! \leq \!\chi^2_{\textrm{crit},\alpha})\! = \! (1-\!\alpha)$, where $P(\chi^2\! \leq\! x)$ is the cumulative distribution function for a $\chi^2$-distribution with $k\! =\! p$ degrees of freedom.\footnote{When estimating the best fit entropy exponent and local Hausdorff dimension, we were only interested in one out the $p$ fit parameters. In this case, we should match to a $\chi^2$-distribution with one degree of freedom.} We can determine $\chi^2_{\textrm{crit},\alpha}$ by the use of quantile functions or lookup tables. The $(1\! -\! \alpha)\%$ confidence interval for the fit parameters $\vthet$ is then defined \cite{avni1976energy} as the region for which 
\begin{equation}
	\chi^2(\vthet) - \chi^2_\textrm{min} < \chi^2_{\textrm{crit},\alpha}.
    \label{eq:ellipsoid-ci}
\end{equation}
Typically, this region is an ellipsoid in $\vthet$-space, which can be determined numerically by performing a grid search around $\vthetm$. As an example, when determining the 95\% confidence inter\-vals for a linear model with two parameters, we find $\chi^2_{\textrm{crit},0.05}\!\approx\! 5.991$, and the joint confidence intervals of the two fit parameters are the region in $\mathbb{R}^2$ for which \eqref{eq:ellipsoid-ci} holds.

\subsection{Potential caveats}
We have used the procedure just described to determine the confidence intervals for the best fit parameters in the main text of this work. However, as pointed out earlier, the analysis rests on several assumptions that do not necessarily apply to our models and measurements. An important prerequisite for using the $\chi^2$-distribution is that the measurement errors are Gaussian, otherwise the sum \eqref{eq:chisq} is not a sum of squares of standard normals. We often found a slight degree of skewness in the distribution of our measurement results, potentially invalidating the use of $\chi^2$-methods. However, the skewness factors were always near zero, so that we may still consider the computed bounds of the confidence intervals to be good approximations to their true values.

A second potential issue is that the models we fit in our work are not linear. Both for the minbu sizes and the microscopic Hausdorff dimension, one of the fit parameters appears in the exponent of an independent variable. Although the best fit parameters for such models can still be obtained by minimizing \eqref{eq:chisq}, determining the correct number of degrees of freedom is known to be difficult \cite{andrae2010dos}. This means that we do not know the proper expectation value of \eqref{eq:chisq}, and therefore do not have a reference point to compare our $\chi^2_\textrm{min}$ to. However, not knowing the true number of degrees of freedom has more serious consequences for computing confidence intervals. The number $p$ of fit parameters in our models is small, $p\!  <\!  4$, and choosing a different number of degrees of freedom near zero has a strong effect on the resulting $\chi^2_{\textrm{crit},\alpha}$. This can significantly alter the width of the corresponding confidence interval. For our purposes, we do not consider this to be a major issue. The main goal of our analysis was to check whether our results are consistent with previously known results from the literature, requiring an order-of-magnitude estimate of the confidence interval. This order of magnitude is not affected if we over- or underestimate the degree-of-freedom count by a few units. 
Furthermore, since all confidence intervals in this work were obtained by using the same methods, any comparison {\it among} our confidence intervals is likely to 
still be meaningful.

\end{appendices}

\vspace{1cm}

\printbibliography

\end{document}